\documentclass{ws-ijmpa}

\usepackage{amsthm}
\usepackage{amsfonts}
\usepackage{xspace}

\usepackage{graphicx}
\usepackage{multirow}
\usepackage{subfigure}
\usepackage{lineno}
\usepackage{hyperref}
\usepackage[capitalise]{cleveref}
\usepackage[normalem]{ulem}
\hypersetup{
	colorlinks=true,
	linkcolor=blue,
	citecolor=blue,
	urlcolor=blue
}
\newcommand{\customUL}{\bgroup\markoverwith{\textcolor{blue}{\rule[-0.3ex]{2pt}{0.6pt}}}\ULon}

\usepackage[super,compress]{cite}
\usepackage{graphicx}

\def\nol {$\ensuremath{0\ell$\xspace}}
\def\l {$\ensuremath{1\ell$\xspace}}
\def\llSC {$\ensuremath{2\ell^\text{ss}$\xspace}}
\def\lll {$\ensuremath{3\ell}$\xspace}
\def\llll {$\ensuremath{4\ell}$\xspace}

\def \sbsbModel{$\ensuremath{\tilde{b}_{1} \rightarrow t W \tilde{\chi}_{1}^{0}$\xspace}}

\newcommand{\pt}{\ensuremath{p_{T}}\xspace}

\newcommand{\met}{\ensuremath{E_\mathrm{T}^\mathrm{miss}}\xspace}

\newcommand{\meff}{\ensuremath{m_\mathrm{eff}}\xspace}

\def\antibar#1{\ensuremath{#1\bar{#1}}}

\def\ttbar{\antibar{t}}

\def\sbottom{\ensuremath{\tilde{b}}}
\def\sbottomone{\ensuremath{\tilde{b}_1}}

\def\ggino{\ensuremath{\mathchoice%
      {\displaystyle\raise.4ex\hbox{$\displaystyle\tilde\chi$}}%
         {\textstyle\raise.4ex\hbox{$\textstyle\tilde\chi$}}%
       {\scriptstyle\raise.3ex\hbox{$\scriptstyle\tilde\chi$}}%
 {\scriptscriptstyle\raise.3ex\hbox{$\scriptscriptstyle\tilde\chi$}}}}

\def\chinop{\ensuremath{\mathchoice%
      {\displaystyle\raise.4ex\hbox{$\displaystyle\tilde\chi^+$}}%
         {\textstyle\raise.4ex\hbox{$\textstyle\tilde\chi^+$}}%
       {\scriptstyle\raise.3ex\hbox{$\scriptstyle\tilde\chi^+$}}%
 {\scriptscriptstyle\raise.3ex\hbox{$\scriptscriptstyle\tilde\chi^+$}}}}
\def\chinom{\ensuremath{\mathchoice%
      {\displaystyle\raise.4ex\hbox{$\displaystyle\tilde\chi^-$}}%
         {\textstyle\raise.4ex\hbox{$\textstyle\tilde\chi^-$}}%
       {\scriptstyle\raise.3ex\hbox{$\scriptstyle\tilde\chi^-$}}%
 {\scriptscriptstyle\raise.3ex\hbox{$\scriptscriptstyle\tilde\chi^-$}}}}
\def\chinopm{\ensuremath{\mathchoice%
      {\displaystyle\raise.4ex\hbox{$\displaystyle\tilde\chi^\pm$}}%
         {\textstyle\raise.4ex\hbox{$\textstyle\tilde\chi^\pm$}}%
       {\scriptstyle\raise.3ex\hbox{$\scriptstyle\tilde\chi^\pm$}}%
 {\scriptscriptstyle\raise.3ex\hbox{$\scriptscriptstyle\tilde\chi^\pm$}}}}
\def\chinomp{\ensuremath{\mathchoice%
      {\displaystyle\raise.4ex\hbox{$\displaystyle\tilde\chi^\mp$}}%
         {\textstyle\raise.4ex\hbox{$\textstyle\tilde\chi^\mp$}}%
       {\scriptstyle\raise.3ex\hbox{$\scriptstyle\tilde\chi^\mp$}}%
 {\scriptscriptstyle\raise.3ex\hbox{$\scriptscriptstyle\tilde\chi^\mp$}}}}

\def\chinoonep{\ensuremath{\mathchoice%
      {\displaystyle\raise.4ex\hbox{$\displaystyle\tilde\chi^+_1$}}%
         {\textstyle\raise.4ex\hbox{$\textstyle\tilde\chi^+_1$}}%
       {\scriptstyle\raise.3ex\hbox{$\scriptstyle\tilde\chi^+_1$}}%
 {\scriptscriptstyle\raise.3ex\hbox{$\scriptscriptstyle\tilde\chi^+_1$}}}}
\def\chinoonem{\ensuremath{\mathchoice%
      {\displaystyle\raise.4ex\hbox{$\displaystyle\tilde\chi^-_1$}}%
         {\textstyle\raise.4ex\hbox{$\textstyle\tilde\chi^-_1$}}%
       {\scriptstyle\raise.3ex\hbox{$\scriptstyle\tilde\chi^-_1$}}%
 {\scriptscriptstyle\raise.3ex\hbox{$\scriptscriptstyle\tilde\chi^-_1$}}}}
\def\chinoonepm{\ensuremath{\mathchoice%
      {\displaystyle\raise.4ex\hbox{$\displaystyle\tilde\chi^\pm_1$}}%
         {\textstyle\raise.4ex\hbox{$\textstyle\tilde\chi^\pm_1$}}%
       {\scriptstyle\raise.3ex\hbox{$\scriptstyle\tilde\chi^\pm_1$}}%
 {\scriptscriptstyle\raise.3ex\hbox{$\scriptscriptstyle\tilde\chi^\pm_1$}}}}

\def\chinotwop{\ensuremath{\mathchoice%
      {\displaystyle\raise.4ex\hbox{$\displaystyle\tilde\chi^+_2$}}%
         {\textstyle\raise.4ex\hbox{$\textstyle\tilde\chi^+_2$}}%
       {\scriptstyle\raise.3ex\hbox{$\scriptstyle\tilde\chi^+_2$}}%
 {\scriptscriptstyle\raise.3ex\hbox{$\scriptscriptstyle\tilde\chi^+_2$}}}}
\def\chinotwom{\ensuremath{\mathchoice%
      {\displaystyle\raise.4ex\hbox{$\displaystyle\tilde\chi^-_2$}}%
         {\textstyle\raise.4ex\hbox{$\textstyle\tilde\chi^-_2$}}%
       {\scriptstyle\raise.3ex\hbox{$\scriptstyle\tilde\chi^-_2$}}%
 {\scriptscriptstyle\raise.3ex\hbox{$\scriptscriptstyle\tilde\chi^-_2$}}}}
\def\chinotwopm{\ensuremath{\mathchoice%
      {\displaystyle\raise.4ex\hbox{$\displaystyle\tilde\chi^\pm_2$}}%
         {\textstyle\raise.4ex\hbox{$\textstyle\tilde\chi^\pm_2$}}%
       {\scriptstyle\raise.3ex\hbox{$\scriptstyle\tilde\chi^\pm_2$}}%
 {\scriptscriptstyle\raise.3ex\hbox{$\scriptscriptstyle\tilde\chi^\pm_2$}}}}

\def\nino{\ensuremath{\mathchoice%
      {\displaystyle\raise.4ex\hbox{$\displaystyle\tilde\chi^0$}}%
         {\textstyle\raise.4ex\hbox{$\textstyle\tilde\chi^0$}}%
       {\scriptstyle\raise.3ex\hbox{$\scriptstyle\tilde\chi^0$}}%
 {\scriptscriptstyle\raise.3ex\hbox{$\scriptscriptstyle\tilde\chi^0$}}}}

\def\ninoone{\ensuremath{\mathchoice%
      {\displaystyle\raise.4ex\hbox{$\displaystyle\tilde\chi^0_1$}}%
         {\textstyle\raise.4ex\hbox{$\textstyle\tilde\chi^0_1$}}%
       {\scriptstyle\raise.3ex\hbox{$\scriptstyle\tilde\chi^0_1$}}%
 {\scriptscriptstyle\raise.3ex\hbox{$\scriptscriptstyle\tilde\chi^0_1$}}}}
\def\ninotwo{\ensuremath{\mathchoice%
      {\displaystyle\raise.4ex\hbox{$\displaystyle\tilde\chi^0_2$}}%
         {\textstyle\raise.4ex\hbox{$\textstyle\tilde\chi^0_2$}}%
       {\scriptstyle\raise.3ex\hbox{$\scriptstyle\tilde\chi^0_2$}}%
 {\scriptscriptstyle\raise.3ex\hbox{$\scriptscriptstyle\tilde\chi^0_2$}}}}
\def\ninothree{\ensuremath{\mathchoice%
      {\displaystyle\raise.4ex\hbox{$\displaystyle\tilde\chi^0_3$}}%
         {\textstyle\raise.4ex\hbox{$\textstyle\tilde\chi^0_3$}}%
       {\scriptstyle\raise.3ex\hbox{$\scriptstyle\tilde\chi^0_3$}}%
 {\scriptscriptstyle\raise.3ex\hbox{$\scriptscriptstyle\tilde\chi^0_3$}}}}
\def\ninofour{\ensuremath{\mathchoice%
      {\displaystyle\raise.4ex\hbox{$\displaystyle\tilde\chi^0_4$}}%
         {\textstyle\raise.4ex\hbox{$\textstyle\tilde\chi^0_4$}}%
       {\scriptstyle\raise.3ex\hbox{$\scriptstyle\tilde\chi^0_4$}}%
 {\scriptscriptstyle\raise.3ex\hbox{$\scriptscriptstyle\tilde\chi^0_4$}}}}

\def\sbottom{\ensuremath{\tilde{b}}}
\def\sbottomone{\ensuremath{\tilde{b}_1}}

\graphicspath{{Figures/}}

\begin{document}
\markboth{Otilia Ducu}{Experimental search potential for \sbsbModel via $\tilde{\chi}^{\pm}_{1}$, with $\geq2 \ell^\text{ss}$}

%
%

\title{Experimental search potential for sbottom via $\tilde\chi^{\pm}_1$ decays at the LHC Run-3 and HL-LHC, in final states with same-sign leptons and multiple jets
}

\author{Otilia Ducu}

\address{Horia Hulubei National Institute of Physics and Nuclear Engineering (IFIN-HH)\\
Magurele, Ilfov, Romania (077125)\\
otilia.ducu@gmail.com}

\maketitle


\begin{abstract}

This paper explores the experimental search potential for sbottom pair production in an R-parity conserving scenario at the LHC Run-3 and HL-LHC. The sbottom decays with a 100\% BR via a chargino, $\tilde{b}_1 \to t \tilde{\chi}_1^\pm$, which subsequently decays to a $W$ boson and a neutralino, $\tilde{\chi}_1^\pm \to W \tilde{\chi}_1^0$, also with a 100\% BR. The study follows the ATLAS object definitions and event selection criteria from Ref.~\citen{ATLAS:2019fag}, focusing on Rpc2L1b and Rpc2L2b signal regions defined with same-sign leptons and at least one $b$-tagged jet. Projected exclusion limits are presented in the $\tilde{b}_1$ - $\tilde{\chi}_1^0$ mass plane for three center-of-mass energies (13~TeV, 13.6~TeV, and 14~TeV) and three integrated luminosity scenarios (139~fb$^{-1}$, 300~fb$^{-1}$, and 3000~fb$^{-1}$).

\end{abstract}



%
%

\section{Introduction}	
\label{sec:intro}

Supersymmetry~\cite{Martin:1997ns}, or simply SUSY, is one of the most well-known and preferred extensions of the Standard Model (SM) of particle physics. This preference is largely due to the solutions SUSY provides to the SM gauge hierarchy problem, achieved without requiring excessive fine-tuning of fundamental parameters by predicting superpartners for each SM particle. The lightest SUSY particle (LSP) is also considered a good dark matter candidate. Additionally, SUSY predicts a perfect unification of the strong, weak, and electromagnetic interactions at the Planck scale.

\begin{figure}[!th]
\begin{center}
	\subfigure[\label{fig:SUSY_xsec}]{\includegraphics[width=0.66\columnwidth]{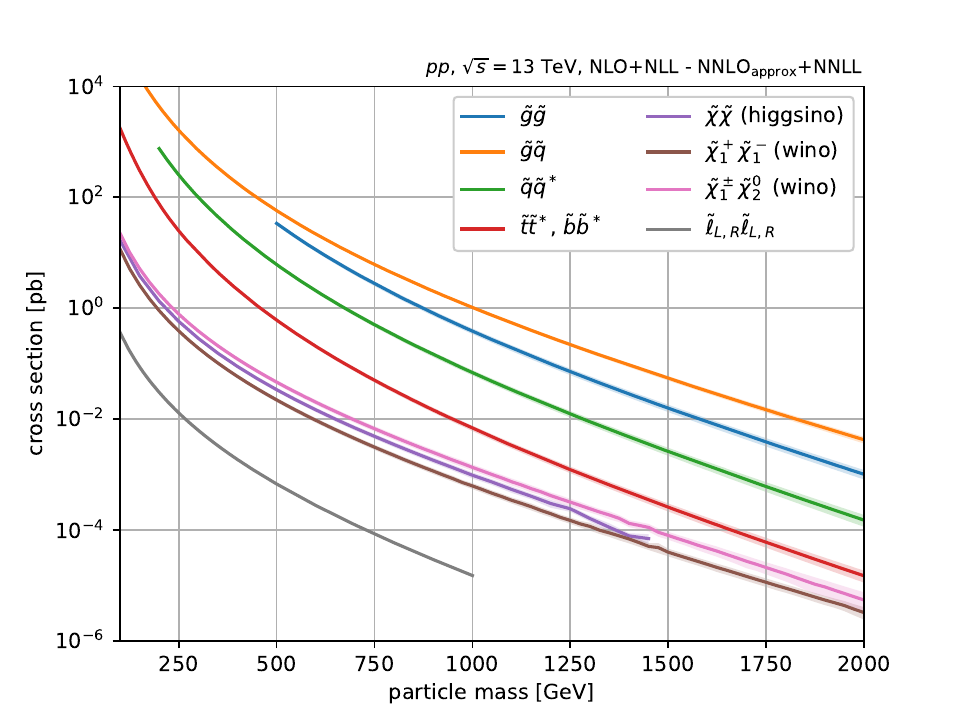}}
	\subfigure[\label{fig:sbottom_Diag}]{\includegraphics[width=0.3\columnwidth]{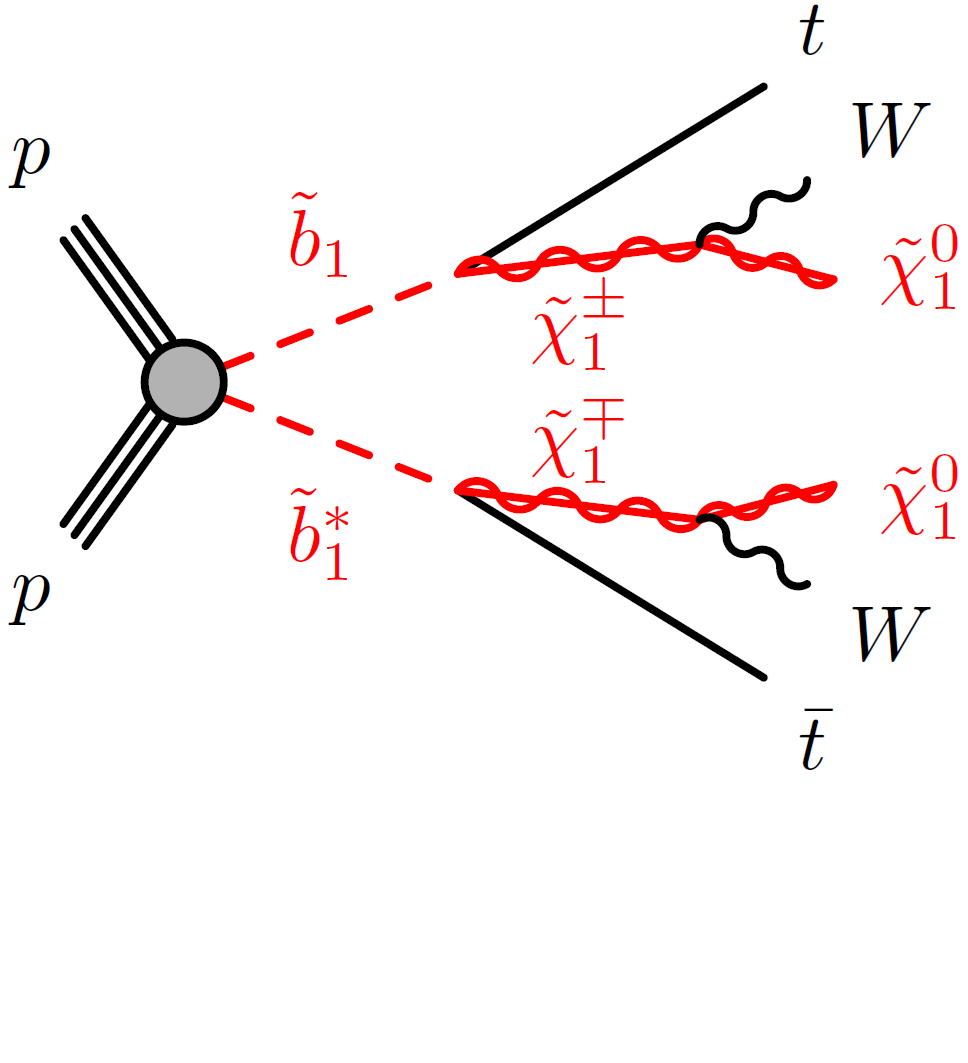}}
	\vspace{-0.3cm}	
\end{center}
\caption{ 
	(a) Cross section predictions for several SUSY production processes at $\sqrt s = 13 TeV$, as determined by the LHC SUSY Cross Section Working Group~\cite{ATLAS:XSec,Borschensky:2014cia}. (b) Diagram representative for the sbottom pair production model considered in this paper.
}
\end{figure}

At the LHC~\cite{Evans:2008zzb}, both the ATLAS~\cite{ATLAS:2008xda} and CMS~\cite{CMS:2008xjf} experiments have dedicated SUSY physics programs. These programs cover a wide range of SUSY production modes: gluino and squark direct production, $3^{rd}$ generation squark direct production, and electroweakinos direct production~\cite{ATLAS:Susy,CMS:Susy}. Among these production modes, the highest production cross-section is for the gluinos and squarks, followed closely by the $3^{rd}$ generation squark production, as illustrated in \cref{fig:SUSY_xsec}.

Of all the SUSY simplified models studied at the LHC, this paper focuses on the sbottom pair production in an R-parity~\cite{FARRAR1978575} conserving scenario. As illustrated in \cref{fig:sbottom_Diag}, the sbottom undergoes a one-step decay via a chargino, $\sbottom_1 \to t \chinoonepm$ (100\% BR), with the chargino decaying as $\chinoonepm \to W \ninoone$ (100\% BR). The sbottom mass is varied from 600 to 1700~GeV, while the $\ninoone$ LSP mass ranges from 50 to 1425~GeV. As in ATLAS Ref.~\citen{ATLAS:2019fag}, the chargino mass is set to be 100~GeV more than the LSP mass, ensuring always on-shell $W$ bosons in its decay. Other LSP mass choices, such as 50 GeV as in CMS Ref.~\citen{CMS:2020cpy}, are also possible but not considered in this paper. For the decay to occur, the sbottom mass must be greater than the sum of the SM top quark mass and the chargino mass: $m_{\sbottom_1} > 172.76 \text{ GeV} + m_{\chinoonepm}$. The masses of the gluinos and 1$^{st}$ and $2^{nd}$ generation squarks are decoupled (set to 4.5~TeV). 

To perform the analysis presented in this paper, signal samples for the sbottom pair production model are generated from leading order matrix elements with up to two extra partons, using the \texttt{MadGraph}~\cite{Alwall:2014hca} generator version \verb|MG5_aMC_v3.5.5|. These are  interfaced with \texttt{Pythia~8}~\cite{Bierlich:2022pfr,Sjostrand:2014zea} for the modeling of the SUSY decay chain, showering, and hadronization. Jet-parton matching  follows the MLM~\cite{Pytha_MLM-Merging} prescription, with a matching scale
set to one quarter of the sbottom mass. 
MC sample generations are done at the center-of-mass energy $\sqrt{s}$ of 13~TeV (LHC Run-2), 13.6~TeV (LHC Run-3), and 14~TeV (probably at HL-LHC).

The signal event samples are processed using the \texttt{DELPHES}~\cite{deFavereau:2013fsa} framework for a fast and realistic simulation of the ATLAS detector. The ATLAS parameter card from \texttt{DELPHES} is used, with several modifications to best match the object selections and definitions from ATLAS Ref.~\citen{ATLAS:2019fag}. Specifically, for jets the anti-$k_\mathrm{T}$ algorithm~\cite{Cacciari:2008gp} with a radius $R = 0.4$ is used instead of $R = 0.6$. The selection efficiency for $b$-tagged jets is changed from 80\% to 70\%. The selection efficiencies for electrons and muons are updated based on the latest publications (see Refs.~\citen{ATLAS:2023dxj,ATLAS:2020auj}), and the isolation identification working points are revised as per ATLAS Ref.~\citen{ATLAS:2019fag}. The two leading leptons (ordered by their transverse momentum, \pt) should satisfy the $\pt > 20$~GeV condition. In addition, only additonal leptons that meet the criteria of \pt~$>10$~GeV and $|\eta|<2.47$ for electrons and $|\eta|<2.5$ for muons are considered.

To analyze the signal samples, the \texttt{SimpleAnalysis}~\cite{ATLAS:2022yru} framework is used. The analysis described in ATLAS Ref.~\citen{ATLAS:2019fag} is implemented here to obtain event counts at different selection steps and calculate the region acceptance $A$\footnote{The acceptance $A$ is calculated as the ratio of the number of events passing the region definition to the total number of events in the sample.}. The statistical significance $Z$ of the signal within the defined signal regions is also computed. Each event has a weight attributed to account for the \texttt{MadGraph} generator weight, production cross-section, and the ATLAS total integrated luminosity: 139~fb$^{-1}$ (obtained at the end of LHC Run-2), 300~fb$^{-1}$ (could be obtained at the end of LHC Run-3), and 3000~fb$^{-1}$ (possible to be achieved at HL-LHC). The production cross-section is taken from Ref.~\citen{ATLAS:XSec}.

\begin{equation}
	Z = \pm\sqrt{2} \times \sqrt{ n \mathrm{ln} \frac{n(b+\sigma^2)}{b^2+n\sigma^2} - \frac{b^2}{\sigma^2}\mathrm{ln}\frac{b^2+n\sigma^2}{b(b+\sigma^2)}},
	\label{eqn:Zn_function}
\end{equation}
The signal significance $Z$ is computed using the formula presented in \cref{eqn:Zn_function}~\cite{Cowan:2010js}. Here, $n$ is the total number of events, while $b$ is the total number of background events. The $\sigma$  parameter represents the uncertainty on the background. As outlined in Ref.~\citen{Cowan:2010js}, the signal significance $Z$ is a key metric in particle physics for evaluating the rejection of a background hypothesis. A $Z$ value of 1.64 corresponds to a $p$-value of 0.05 at a 95\% confidence level~\cite{Cowan:2010js}, which is generally sufficient to exclude a signal hypothesis. A $Z$ value of 5, on the other hand, equates to a $p$-value of $2.87 \times 10^{-7}$~\cite{Cowan:2010js}, which is a reasonable threshold for declaring a discovery. The studies documented in this paper use these $Z$ values to quantify the exclusion or discovery potential at the LHC and HL-LHC.

\section{ATLAS search for \sbsbModel}
\label{sec:ATLAS_SS3L_Sbottom}

The signal object selections and definitions from the ATLAS Ref.~\citen{ATLAS:2019fag} analysis are used in the study documented in this paper. Given that experimental final states with zero (\nol) or one (\l) lepton, as well as two leptons of opposite charge, are dominated by a high amount of SM background, both studies focus on final states with at least two (\llSC) leptons of the same electric charge (SS), or three (\lll) or four ($\llll$) leptons. Indeed, SM processes like di-jets and vector bosons $V$ ($W$, $Z$) in association with jets have significantly higher production cross-sections compared to processes like $VV$, $\ttbar H$, or $\ttbar V$~\cite{ATLAS:SMSummary}. Even if the $\geq$\llSC selection suffers from low BRs, and these final states are dominated by the fake/non-prompt lepton and electron charge flip backgrounds, regions enriched in signal can successfully be defined by exploiting the kinematic differences between the various background sources and the signal.

\begin{figure}[!th]
	\begin{center}
		\includegraphics[width=0.32\columnwidth]{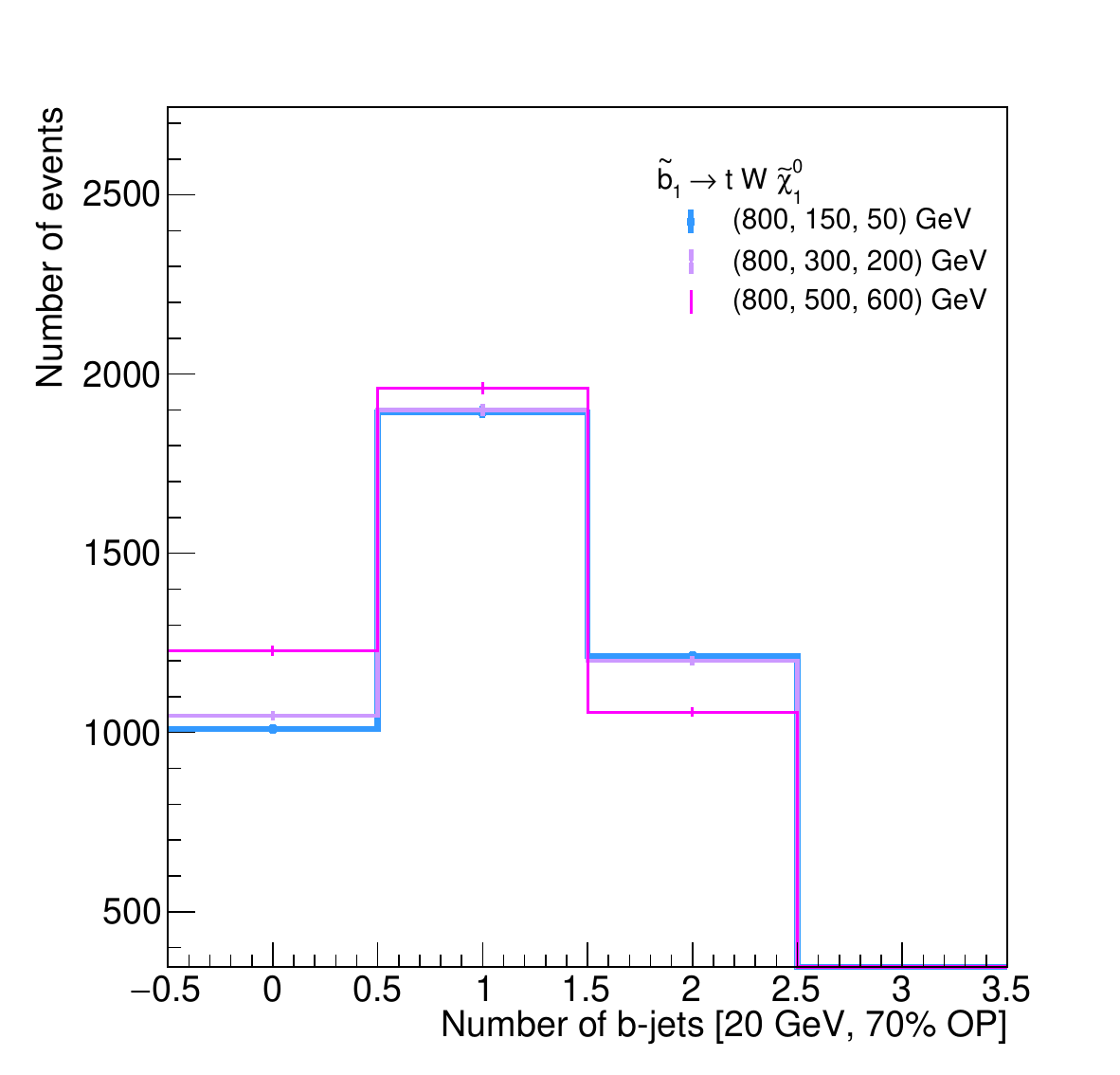}
		\includegraphics[width=0.32\columnwidth]{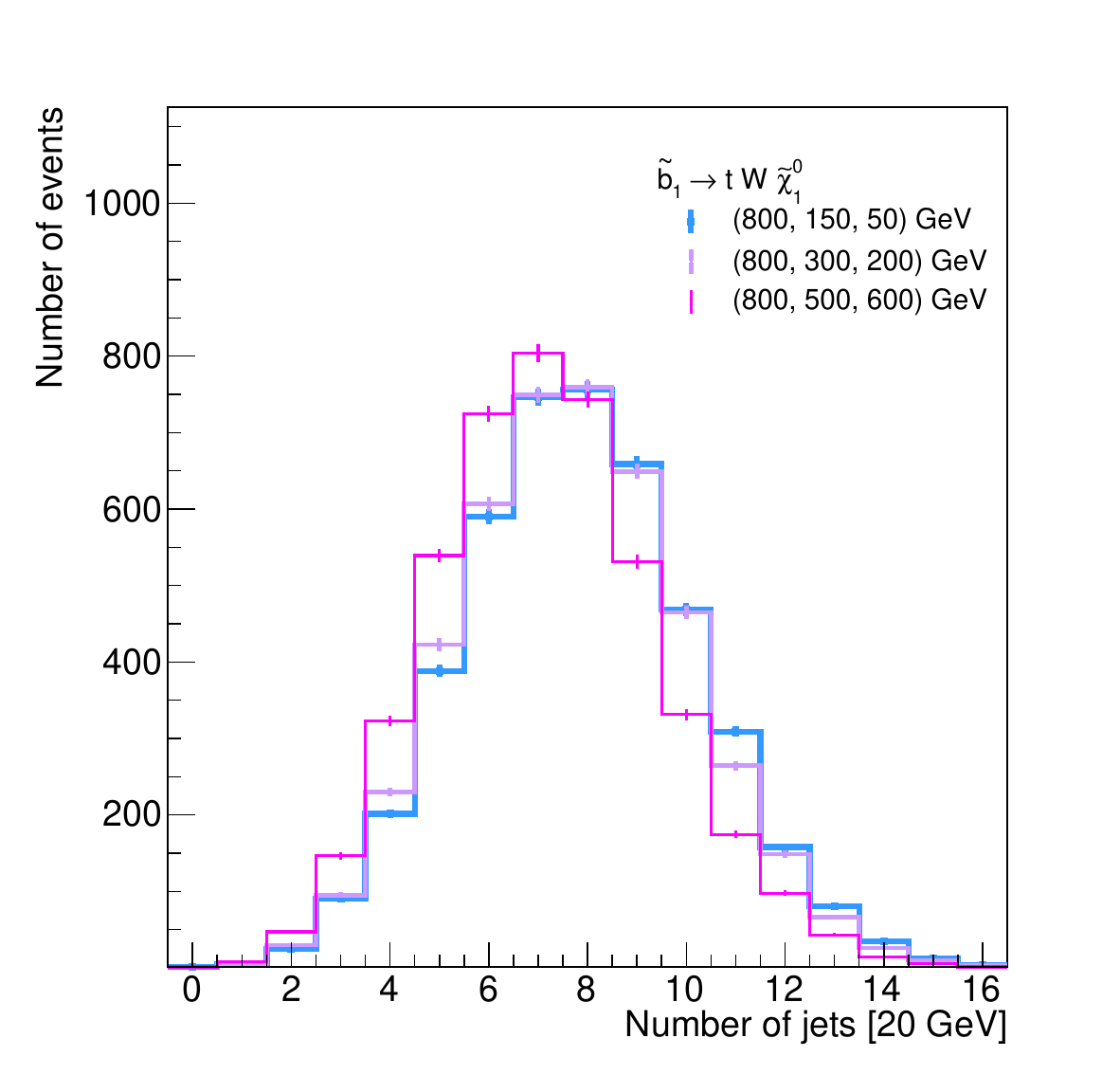}
		\includegraphics[width=0.32\columnwidth]{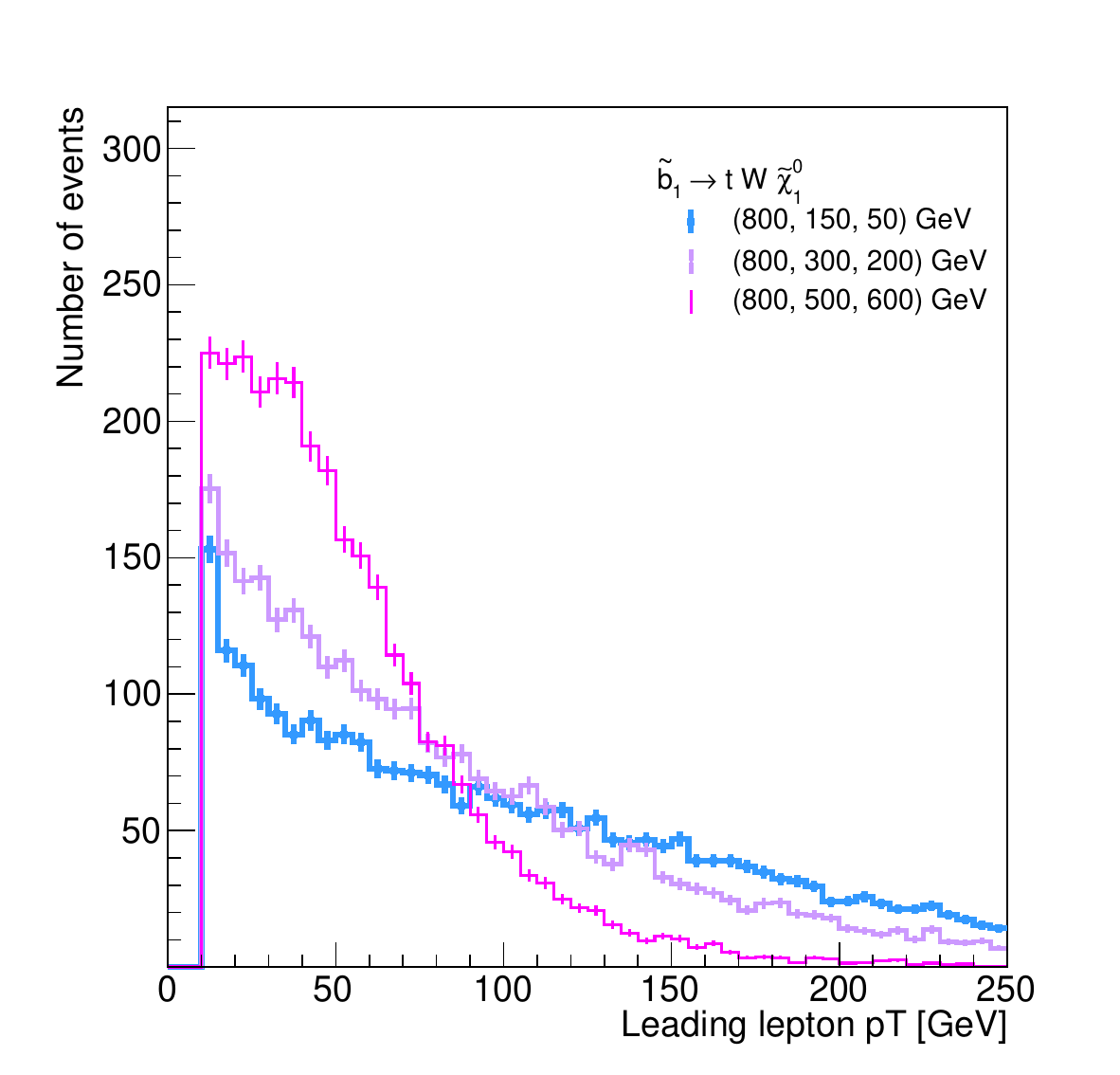}
		\includegraphics[width=0.32\columnwidth]{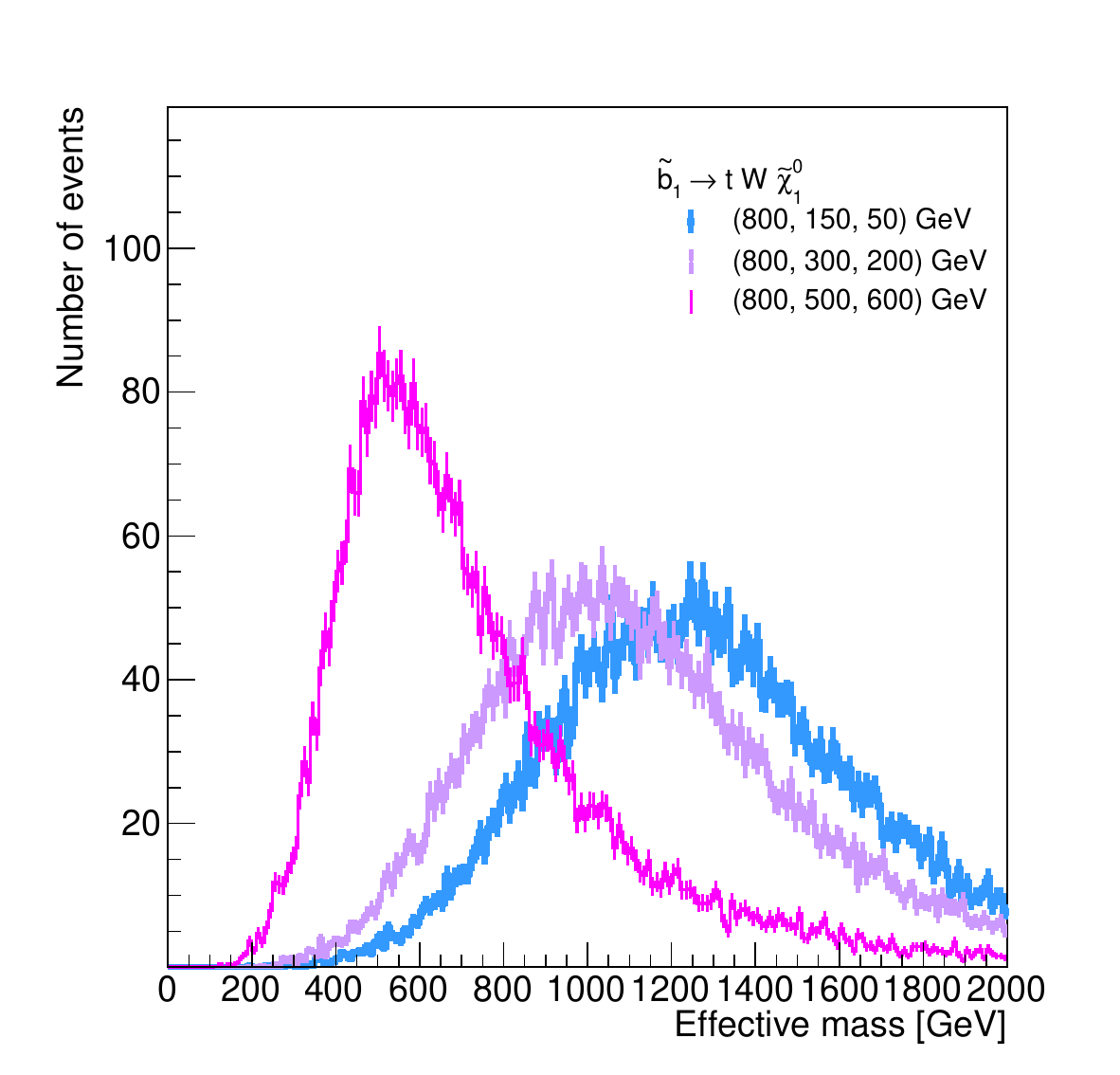}
		\includegraphics[width=0.32\columnwidth]{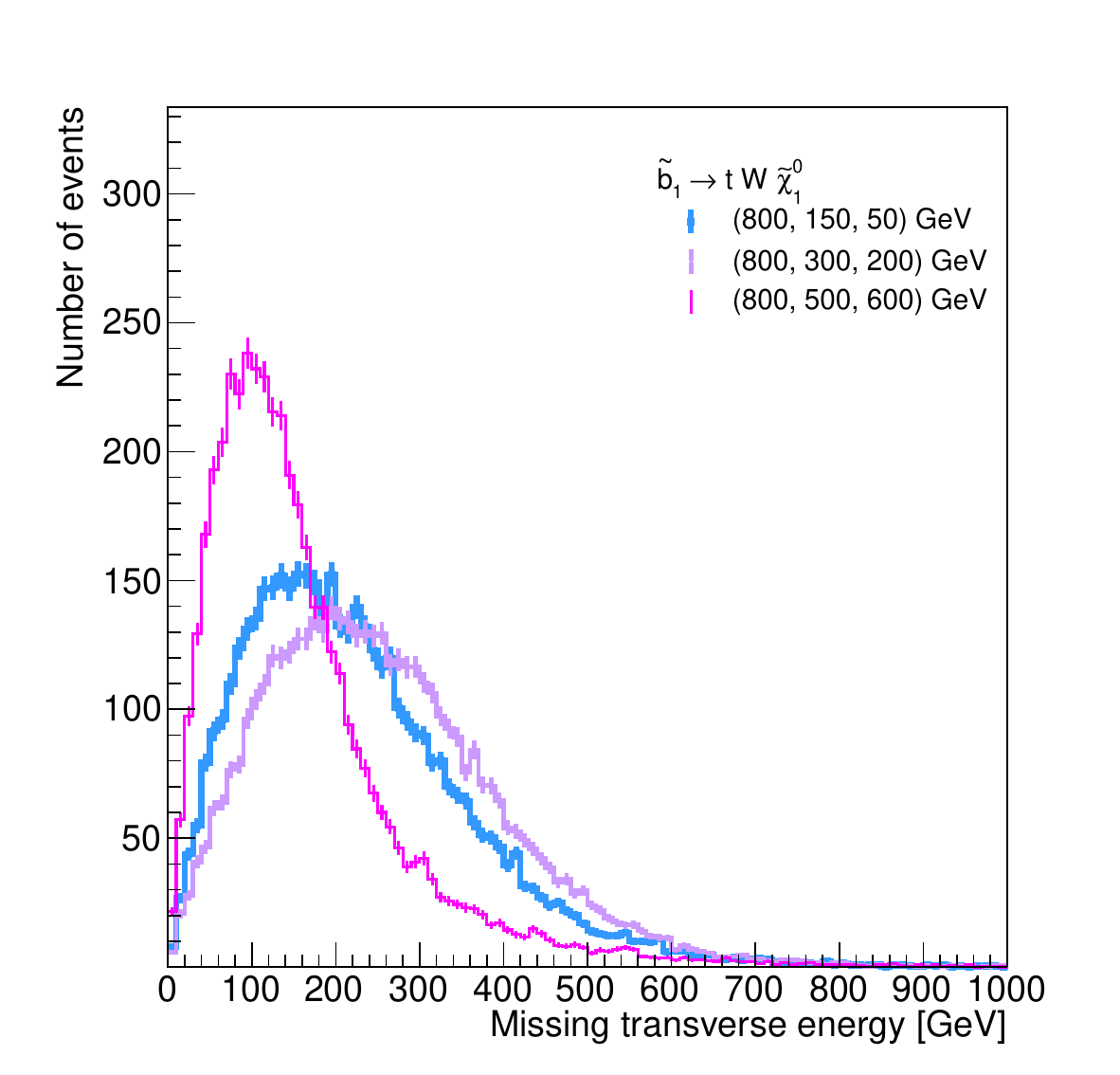}
		\includegraphics[width=0.32\columnwidth]{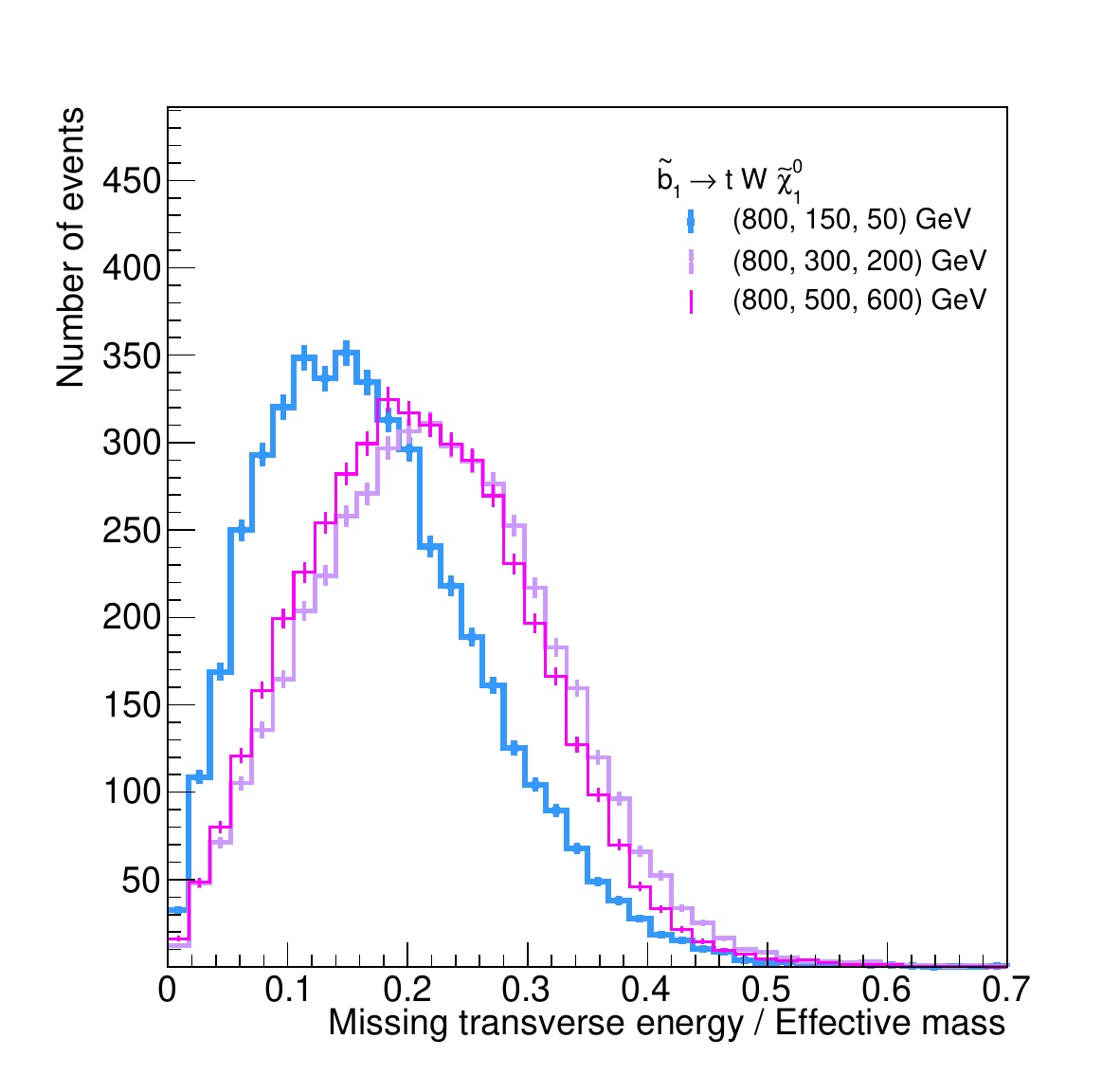}
		\vspace{-0.3cm}	
	\end{center}
	\caption{ 
		Distributions showing the number of ($b$-tagged) jets, the leading lepton $\pt$, \meff, \met, and the \met to \meff ratio in MC signal simulations. The statistical uncertainty is shown with a vertical line, and the distributions are normalized to 139~fb$^{-1}$ ($\sqrt s = 13$~TeV). 
	}
	\label{fig:Distrib}
\end{figure}

In this section, selected results are shown for three benchmark signal mass points, ($\sbottom_1$, $\chinoonepm$, $\ninoone$). The (800, 150, 50)~GeV mass point is representative of the very boosted region, dominated by events with energetic objects, thus high \meff~\footnote{\meff stands for the sum of the signal leptons and jets \pt, and the missing transverse energy}. To cover the compressed region of the phase space, where the mass difference between the $\sbottom_1$ and $\ninoone$ is small, the (800, 500, 600)~GeV mass point is considered. In this region, the objects present in the final state are softer, thus characterized by lower $\pt$, making them harder to detect and distinguish from the background. The \met~\footnote{\met is the magnitude of the missing transverse momentum.} is also low, making signal versus background separation even more challenging. To cover the intermediate region of the SUSY phase space, where objects are neither highly boosted nor in a compressed spectrum, the (800, 300, 200)~GeV mass point is selected. 
\cref{fig:Distrib}, shows distributions of the number of ($b$-tagged) jets, the leading lepton $\pt$, \meff, \met, and the \met to \meff ratio for the three selected benchmark signal points.

\begin{table}[!th]
	\centering
	\tbl{
		Event counting and acceptance after the lepton selection discussed in the text, for the three representative signal ($\sbottom_1$, $\chinoonepm$, $\ninoone$) mass points: 
		(800, 150, 50)~GeV, (800, 300, 200)~GeV and (800, 500, 600)~GeV. The statistical uncertainty is also shown, and the events are normalized to an integrated luminosity of 139~fb$^{-1}$ at $\sqrt{s} = 13$ TeV.
	}
	{\footnotesize
	\def\arraystretch{1.5}
	\setlength{\tabcolsep}{0.0pc}
		\begin{tabular*}{\textwidth}{@{\extracolsep{\fill}}llccc}
			\noalign{\smallskip}\hline\noalign{\smallskip}
			&   & (800, 150, 50) & (800, 300, 200) & (800, 500, 600) \\
			& Selection & N events ($A$) & N events ($A$) & N events ($A$) \\
			\noalign{\smallskip}\hline\noalign{\smallskip}
			& All &                 4533.56 $\pm$ 26.65 (100.00 \%) & 4533.61 $\pm$ 26.72 (100.00 \%) & 4535.85 $\pm$ 26.76 (100.00 \%) \\
			\noalign{\smallskip}\hline\hline\noalign{\smallskip}
			\multirow[c]{6}{*}[0in]{\rotatebox{90}{$\ell$ selection}}
			& $=0\ell$ &             1622.76 $\pm$ 15.95 (35.79 \%)  & 1610.75 $\pm$ 15.93 (35.53 \%) &  1563.17 $\pm$ 15.71 (34.46 \%) \\
			& $=1\ell$ &             1587.34 $\pm$ 15.77 (35.01 \%)  & 1537.98 $\pm$ 15.56 (33.92 \%) &  1434.62 $\pm$ 15.05 (31.63 \%) \\
			& $=2\ell$ &             612.53 $\pm$ 9.80 (13.51 \%)    & 576.82 $\pm$ 9.53 (12.72 \%)   &  518.79 $\pm$ 9.05 (11.44 \%)   \\
			& $=2\ell^{\mathrm{SS}}$ & 213.58 $\pm$ 5.79 (4.71 \%)   & 200.83 $\pm$ 5.62 (4.43 \%)    &  185.25 $\pm$ 5.41 (4.08 \%) \\
			& $=3\ell$ &             187.41 $\pm$ 5.42 (4.13 \%)     & 185.24 $\pm$ 5.40 (4.09 \%)    &  193.30 $\pm$ 5.53 (4.26 \%)\\
			& $\geq4\ell$ &          28.83 $\pm$ 2.13 (0.64 \%)      & 28.04 $\pm$ 2.10 (0.62 \%)     &  34.27 $\pm$ 2.33 (0.76 \%) \\
			\noalign{\smallskip}\hline\hline\noalign{\smallskip}
	\end{tabular*}
	}	
	\label{tab:Sbottom13TeVLepSel}
\end{table}

\cref{tab:Sbottom13TeVLepSel} shows the acceptance $A$, as well as the number of events, after several signal lepton selections: =\nol, =\l, $=2\ell$ with no charge requirement, =\llSC, =\lll and $\geq \llll$. For completeness, the total number of events in the signal MC samples is also shown, in the ``\textit{All}" entry.  The considered integrated luminosity is 139 fb$^{-1}$ ($\sqrt{s} = 13$ TeV). The \nol and \l selections have the highest acceptance for all benchmark mass points, varying between 32\% and 36\%. For the $=2\ell$ selection, the acceptance is smaller, between 11\% and 14\%. Once the \llSC or \lll selection is applied, the acceptance drops to around 4-5\%. As expected, once $\geq \llll$ are requested in the event, the acceptance is very small, less than 1\%. These results also show that the statistics are similar in the selected samples representative of the boosted and intermediate regions of the phase space.

On top of a $\geq$\llSC lepton selection, two signal regions enriched in \sbsbModel potential signals are defined in the ATLAS Ref.~\citen{ATLAS:2019fag} analysis: Rpc2L1b and Rpc2L2b, the latter signal region being defined for better coverage of the very boosted region. As the name suggests, Rpc2L1b (Rpc2L2b) is defined with $\geq 1$ ($\geq 2$) $b$-tagged jets. Even though one would expect higher sensitivity with a two $b$-tagged selection, the 70\% $b$-tagging efficiency~\cite{ATLAS:2019bwq} ensures also a high number of events with one $b$-tagged jets. This is illustrated in the top-left plot of \cref{fig:Distrib}. Both regions are defined with at least 6 jets, but their minimum \pt is 40~GeV and 25~GeV, respectively. Such a high number of jets is well motivated by the various jets produced in the $\sbottom$ decay chain, as shown also in \cref{fig:Distrib} top-middle. A minimum requirement on the $\met$ over $\meff$ ratio of 0.25 is applied in Rpc2L1b, and of 0.14 in Rpc2L2b, respectively. As hinted in \cref{fig:Distrib} bottom-right, the $>0.14$ requirement tends to be too tight. However, according to Ref.~\citen{ATLAS:SS3L_Figs}
this signal region is still found to perform better than Rpc2L1b for some signal mass points at high $\sbottom$ masses. To improve the sensitivity in this region even more, one could consider a binned fit in this $\met$ over $\meff$ variable, or a signal region optimized to cover exactly this very boosted region -- look at Figure~8 a) in Ref.~\citen{ATLAS:2019fag} to see how the exclusion limit seems to be worse for the very boosted region compared to the intermediate region. In Rpc2L2b, $\met > 300$~GeV and $\meff > 1.4$~TeV additional requirements are applied to further remove the background.

\begin{table}[!th]
	\centering
	\tbl{
		Event counting and acceptance after the Rpc2L1b and Rpc2L2b signal regions (pre-)selection discussed in the text, for the three representative signal ($\sbottom_1$, $\chinoonepm$, $\ninoone$) mass points: 
		(800, 150, 50)~GeV, (800, 300, 200)~GeV and (800, 500, 600)~GeV. The statistical uncertainty is also shown, and the events are normalized to an integrated luminosity of 139~fb$^{-1}$ at $\sqrt{s} = 13$ TeV.
	}
	{\footnotesize
	\def\arraystretch{1.5}
	\setlength{\tabcolsep}{0.0pc}
		\begin{tabular*}{\textwidth}{@{\extracolsep{\fill}}llccc}
			\noalign{\smallskip}\hline\noalign{\smallskip}
			&   & (800, 150, 50) & (800, 300, 200) & (800, 500, 600) \\
			& Selection & N events ($A$) & N events ($A$) & N events ($A$) \\
			\noalign{\smallskip}\hline\noalign{\smallskip}
			\multirow[c]{3}{*}[0in]{\rotatebox{90}{Rpc2L1b}}
			 & Pre-sel 1 &           292.24 $\pm$ 6.77 (6.45 \%)     & 282.90 $\pm$ 6.68 (6.24 \%) & 260.11 $\pm$ 6.41 (5.73 \%) \\
			 & Pre-sel 2 &           74.43 $\pm$ 3.42 (1.64 \%)      & 58.12 $\pm$ 3.03 (1.28 \%)  & 30.16 $\pm$ 2.18 (0.67 \%)  \\
			 & {\bf Rpc2L1b} &       8.62 $\pm$ 1.16 (0.19 \%)       & 16.54 $\pm$ 1.61 (0.36 \%)  & 8.84 $\pm$ 1.18 (0.19 \%)   \\
			\noalign{\smallskip}\hline\noalign{\smallskip}
			\multirow[c]{5}{*}[0in]{\rotatebox{90}{Rpc2L2b}}
			 & Pre-sel 1 &           103.58 $\pm$ 4.03 (2.28 \%)     & 96.56 $\pm$ 3.90 (2.13 \%) & 80.54 $\pm$ 3.57 (1.78 \%) \\
			 & Pre-sel 2 &           56.88 $\pm$ 2.99 (1.25 \%)      & 51.82 $\pm$ 2.86 (1.14 \%) & 32.69 $\pm$ 2.27 (0.72 \%) \\
			 & Pre-sel 3 &           13.16 $\pm$ 1.44 (0.29 \%)      & 19.37 $\pm$ 1.75 (0.43 \%) & 6.48 $\pm$ 1.01 (0.14 \%)  \\
			 & Pre-sel 4 &           9.56 $\pm$ 1.22 (0.21 \%)       & 9.45 $\pm$ 1.22 (0.21 \%)  & 3.16 $\pm$ 0.71 (0.07 \%)  \\
			 & {\bf Rpc2L2b} &       9.25 $\pm$ 1.20 (0.20 \%)       & 9.29 $\pm$ 1.21 (0.20 \%)  & 3.16 $\pm$ 0.71 (0.07 \%)  \\

			\noalign{\smallskip}\hline\hline\noalign{\smallskip}

	\end{tabular*}
	}	
	\label{tab:Sbottom13TeVSRSelATLAS}
\end{table}
\begin{figure}[!th]
	\begin{center}\hspace{-0.7cm}
		\includegraphics[width=0.36\columnwidth]{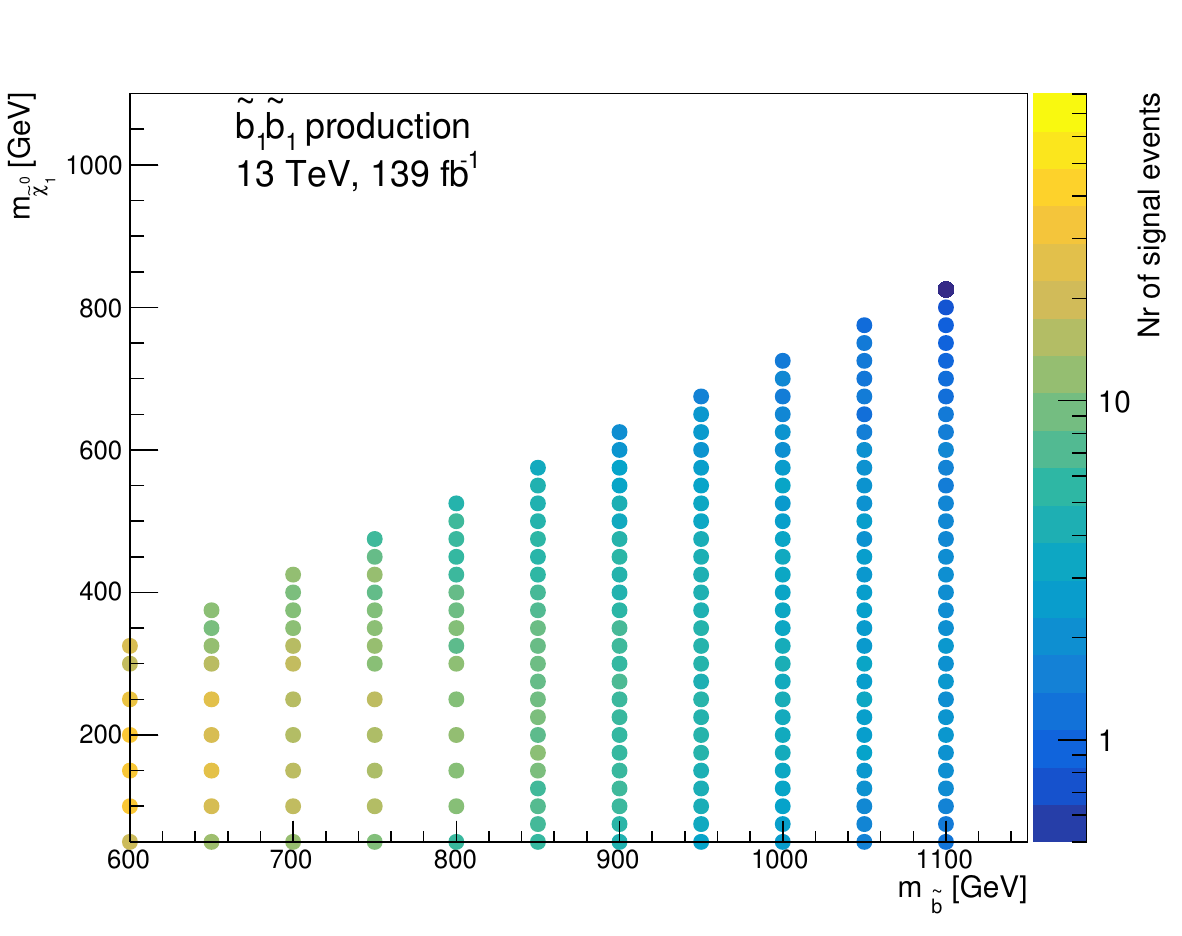}\hspace{-0.3cm}
		\includegraphics[width=0.36\columnwidth]{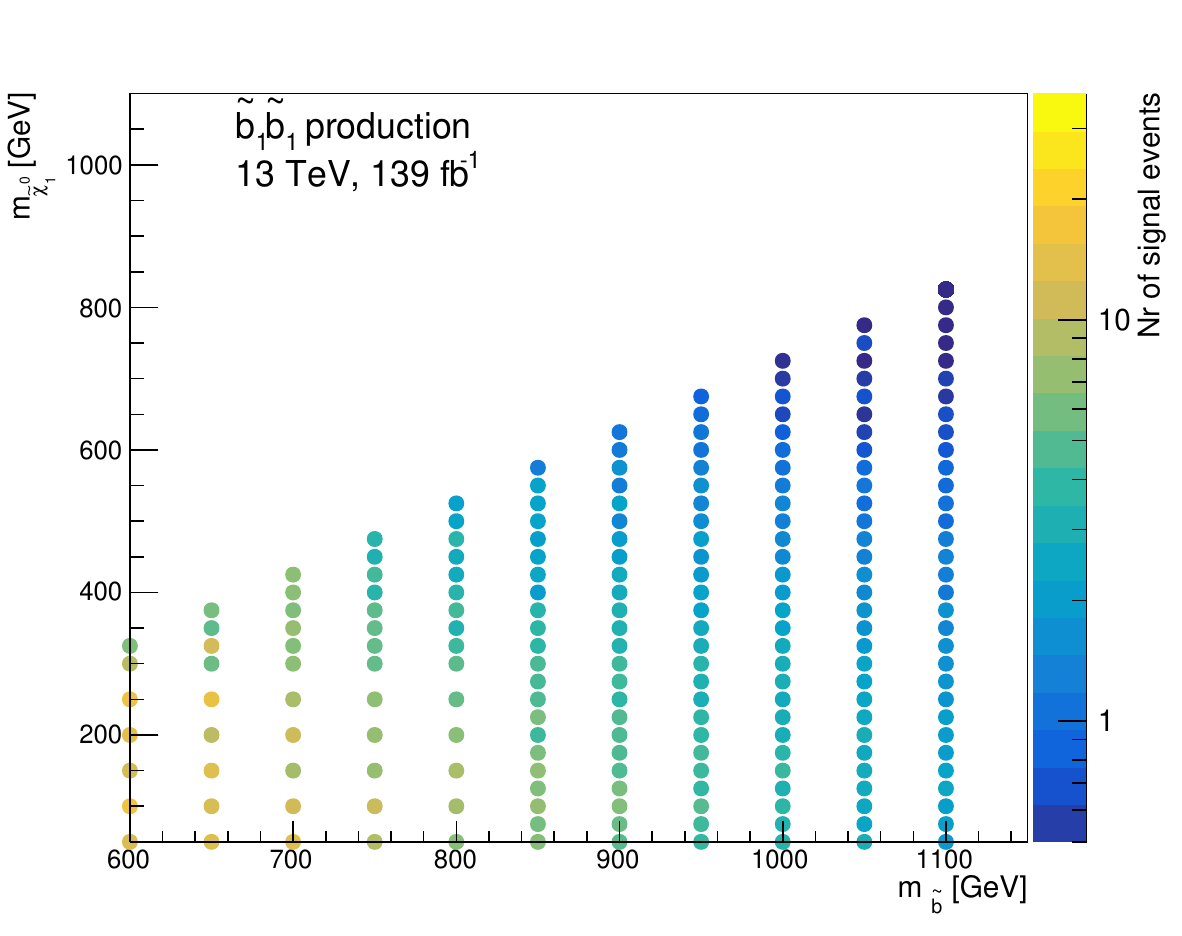}\hspace{-0.3cm}
		\includegraphics[width=0.36\columnwidth]{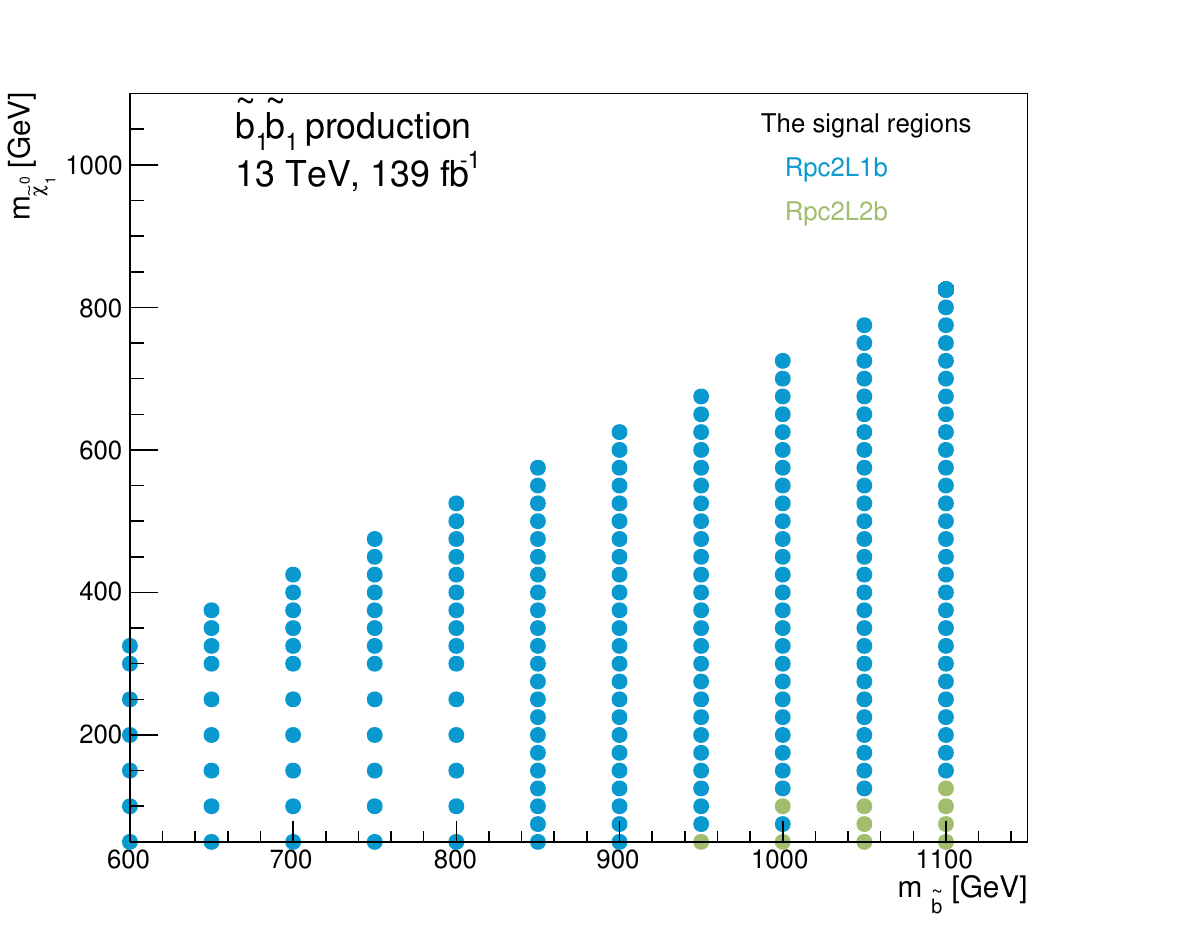}\hspace{-0.7cm}
		\vspace{-0.3cm}
	\end{center}
	\caption{
		Number of events in selected MC \sbsbModel signal samples (see also the $z$ axis), normalized to 139~fb$^{-1}$ ($\sqrt s = 13$~TeV), in Rpc2L1b (left) and Rpc2L2b (middle). The signal region that gives the best signal significance $Z$ for each mass point is also shown (right).
	}
	\label{fig:SignalEvnt_13TeV_ATLAS}
\end{figure}

The impact of the signal region selections on the three selected benchmark points is shown in \cref{tab:Sbottom13TeVSRSelATLAS}. Two and four pre-selection steps are considered for Rpc2L1b and Rpc2L2b, respectively: \vspace{-0.13cm}
\begin{itemize}
	\item Pre-sel 1: $\geq$\llSC, Nr. $b$-tagged jets $\geq 1$,
	\item Pre-sel 2: Pre-sel 1, Nr. jets with $\pt > 40$~GeV $\geq 6$,
	\item {\bf Rpc2L1b}: Pre-sel 2, $\met / \meff > 0.25$.
  \vspace{0.15cm}
	\item Pre-sel 1: $\geq$\llSC, Nr. $b$-tagged jets $\geq 2$,
	\item Pre-sel 2: Pre-sel 1, Nr. jets with $\pt > 25$~GeV $\geq 6$,
	\item Pre-sel 3: Pre-sel 2, $\met > 300$~GeV,
	\item Pre-sel 4: Pre-sel 3, $\meff > 1.4$~TeV,
	\item {\bf Rpc2L2b}: Pre-sel 4, $\met / \meff > 0.14$.	
\end{itemize}\vspace{-0.13cm}
The results show the number of events and the region acceptance. The considered integrated luminosity is 139~fb$^{-1}$ ($\sqrt{s} = 13$ TeV). It can be seen that the Rpc2L1b and Rpc2L2b signal region acceptance is quite small, around 0.2\%. This is expected, as stringent requirements are necessary to remove the large SM and detector backgrounds, as detailed in ATLAS Ref.~\citen{ATLAS:2019fag}. For completeness, the number of signal events, as well as the signal region that gives the best signal significance $Z$ for each mass point, is shown in \cref{fig:SignalEvnt_13TeV_ATLAS} for more signal mass points.

The ATLAS collaboration estimated a total of 6.5$^{+1.5}_{-1.6}$ and 7.8$^{+2.1}_{-2.3}$ background events in Rpc2L1b and Rpc2L2b, respectively. For the 139~fb$^{-1}$ results shown in this paper, these estimations are used, with relative uncertainties of 25\% and 30\%, respectively. For the extrapolations to 300~fb$^{-1}$ and 3000~fb$^{-1}$, the uncertainties were decreased to 20\% and 10\%, as the precision on the total background estimation is expected to improve given the significant increase in the total integrated luminosity. This is reasonable, as the current data-based methods used to estimate backgrounds, like electron charge-flipped or fake/non-prompt leptons, are affected by low statistics~\cite{ATLAS:2019fag}. 

In addition, for the extrapolations to a luminosity of 300~fb$^{-1}$ (3000~fb$^{-1}$), the total number of background events is multiplied by a factor of 2.16 (21.60) to account for the increase in data statistics from 139~fb$^{-1}$. For the projections at 13.6~TeV and 14~TeV center-of-mass energy, the ATLAS Ref.~\citen{ATLAS:2019fag} background estimates are further multiplied by factors of 1.1 and 1.2 to account for the increase in background production cross-sections.

\section{Projected exclusion limits for \sbsbModel model}

\begin{figure}[!th]
	\begin{center}\hspace{-0.9cm}
		\includegraphics[width=0.36\columnwidth]{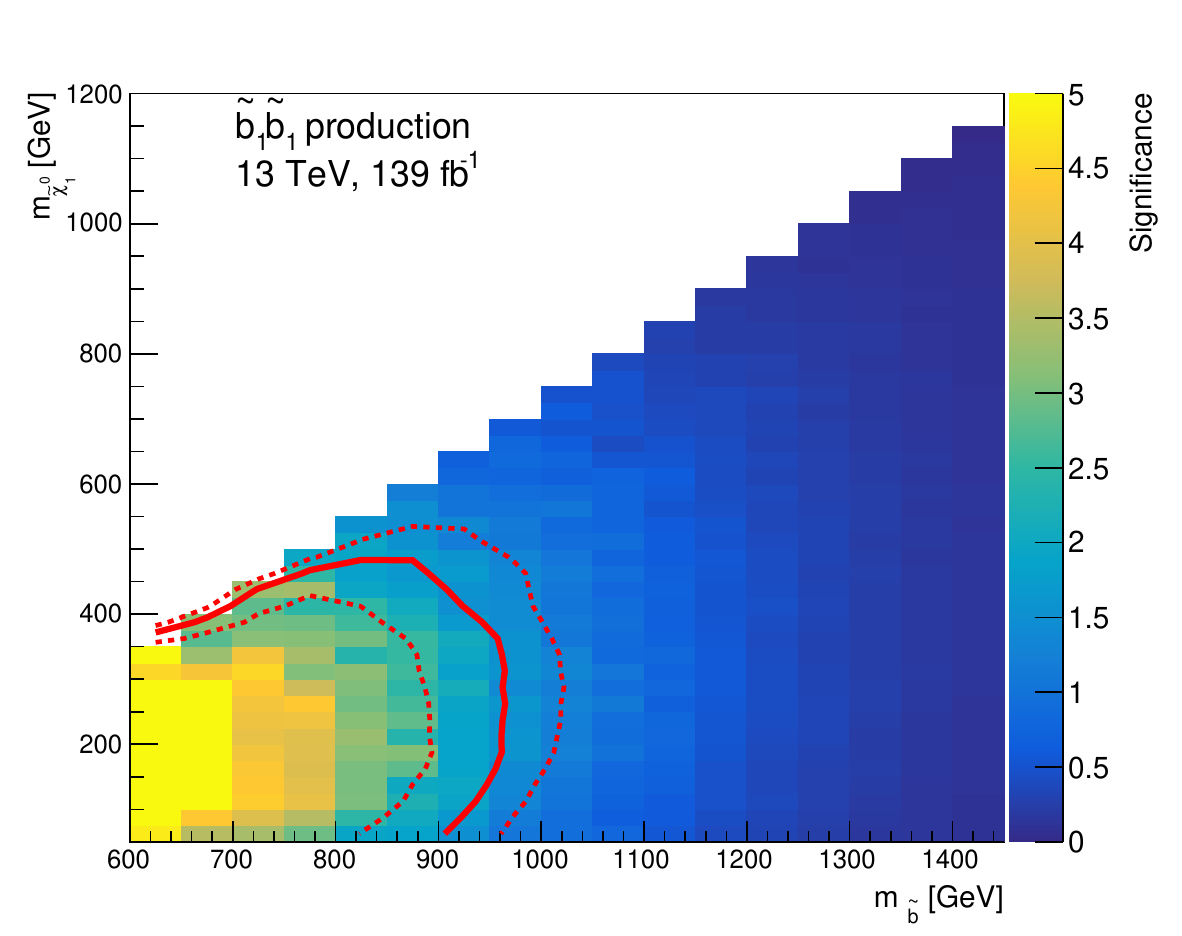}\hspace{-0.35cm}
		\includegraphics[width=0.36\columnwidth]{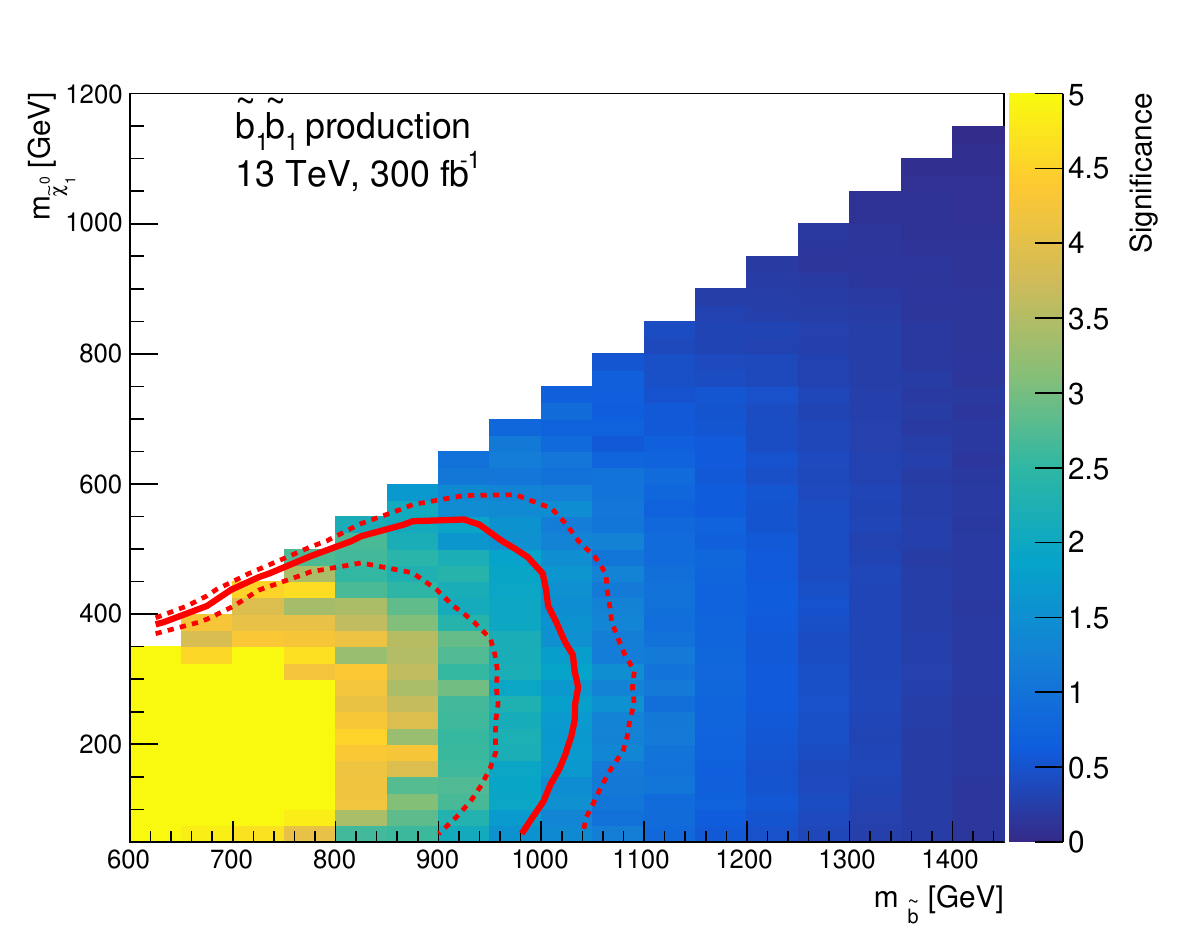}\hspace{-0.35cm}
		\includegraphics[width=0.36\columnwidth]{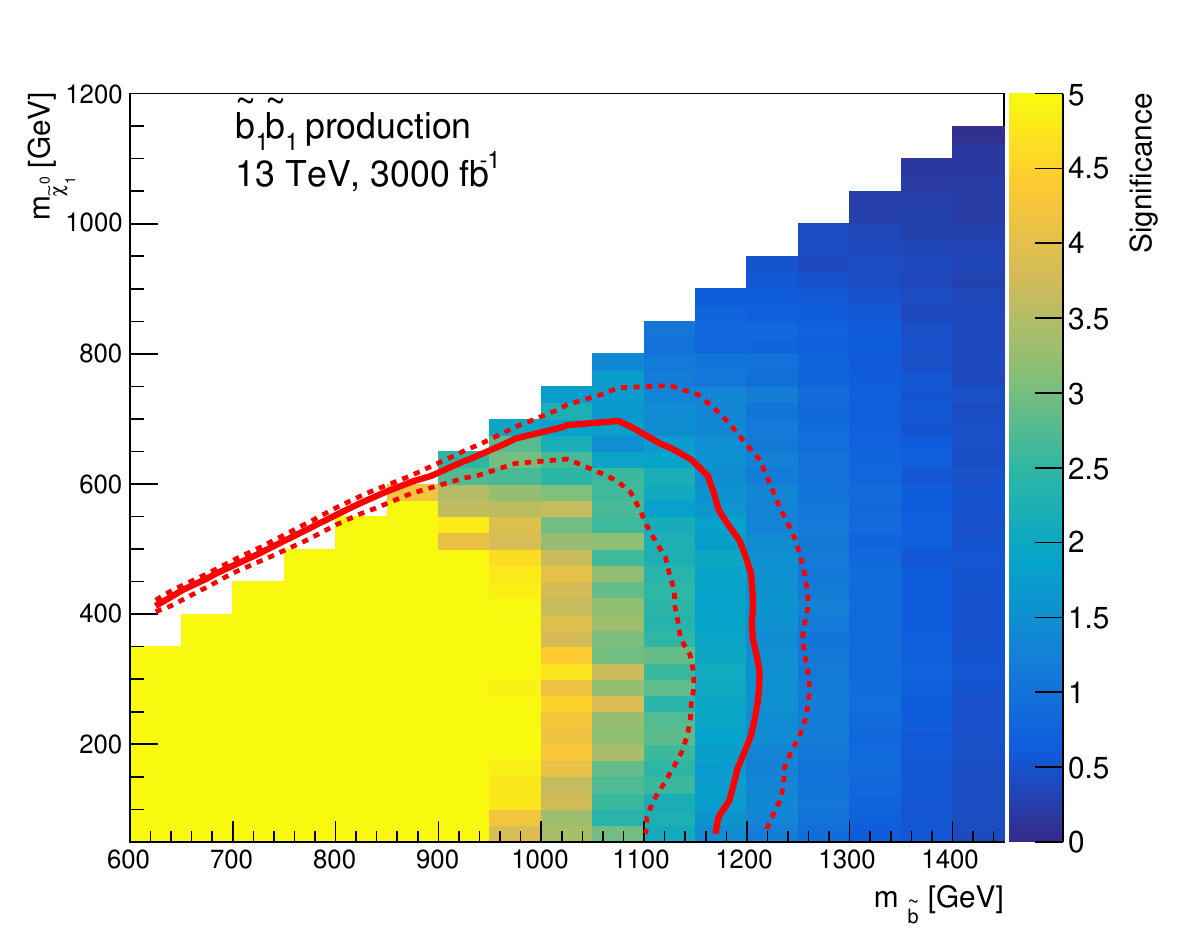}\hspace{-0.7cm}
	\end{center}
	\begin{center}\hspace{-1.3cm}
		\includegraphics[width=0.36\columnwidth]{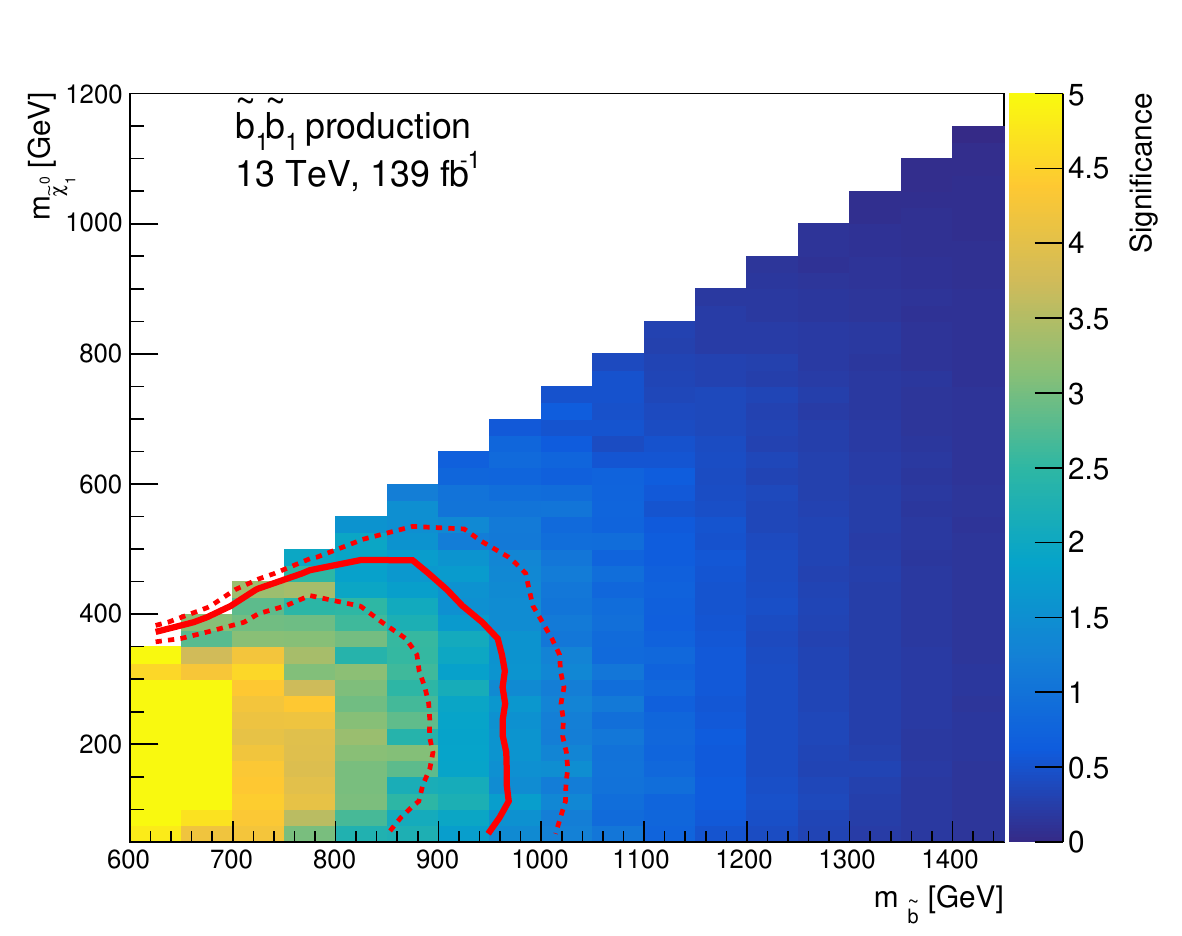}\hspace{-0.35cm}
		\includegraphics[width=0.36\columnwidth]{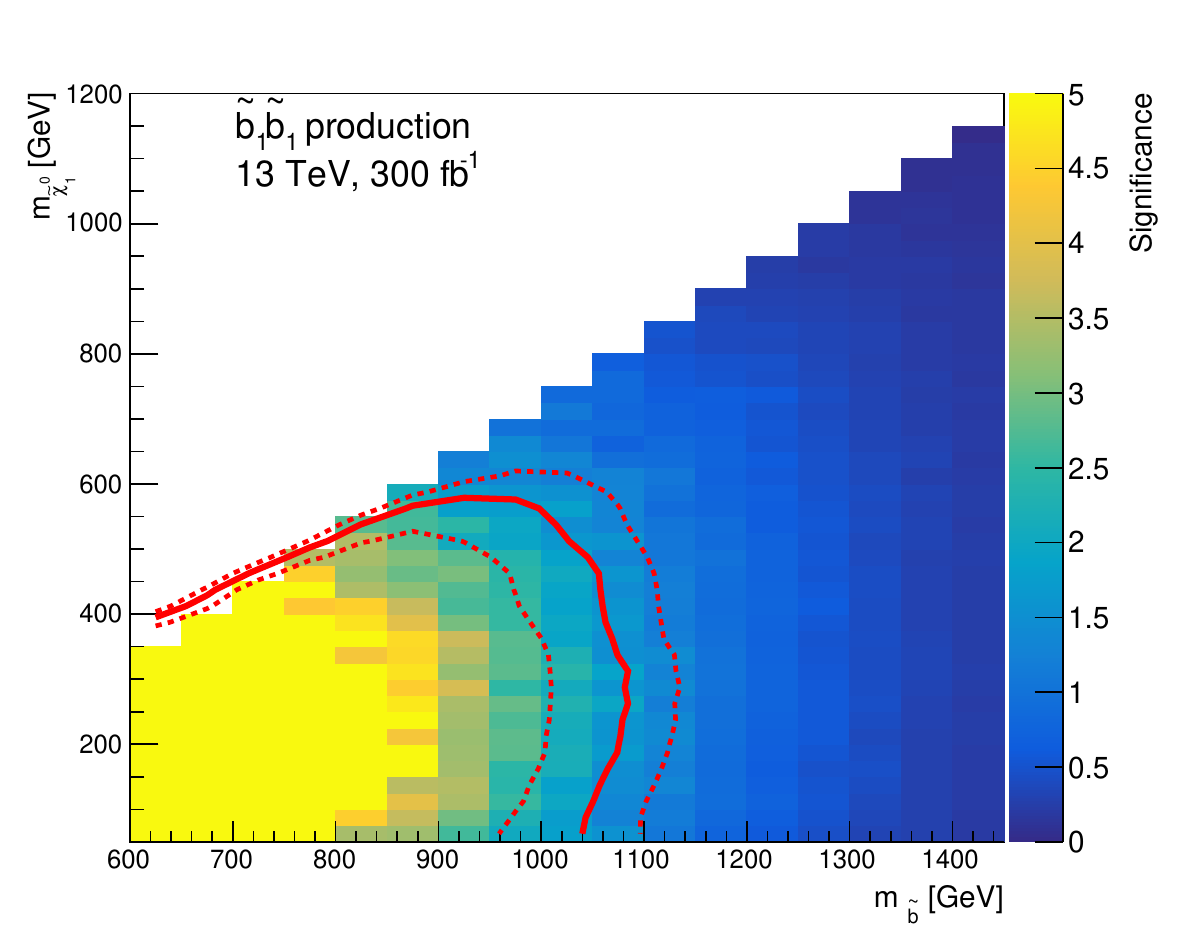}\hspace{-0.35cm}
		\includegraphics[width=0.36\columnwidth]{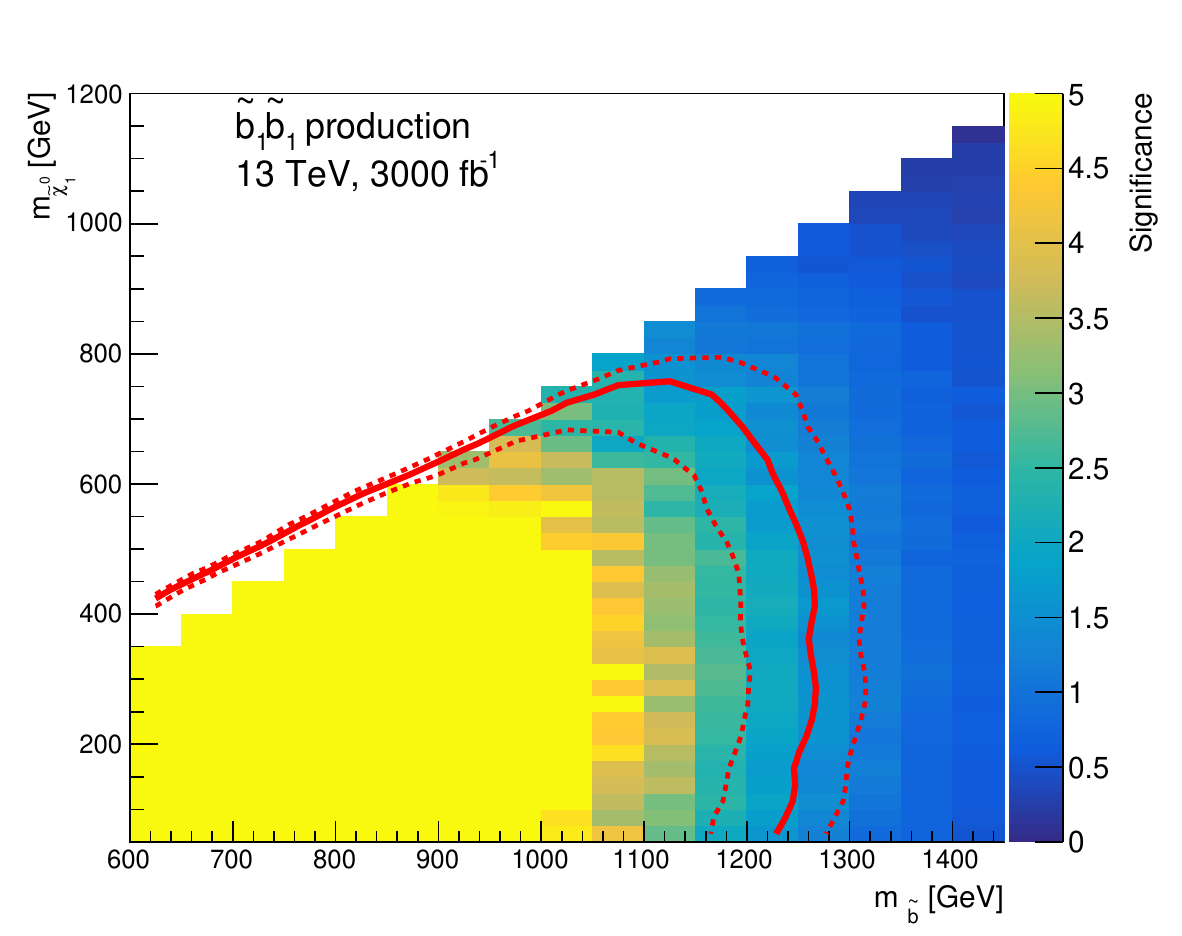}\hspace{-0.7cm}
		\vspace{-0.3cm}
	\end{center}
	\caption{
		Projected exclusion mass limits obtained with the signal region that gives the best signal significance $Z$ for each mass point. Projections are shown for $\sqrt s = 13$~TeV, and total integrated luminosities of 139~fb$^{-1}$ (left), 300~fb$^{-1}$ (middle), and 3000~fb$^{-1}$ (right). 
		The dashed lines correspond to the $\pm 1\sigma$ uncertainty on the signal event yield.
		Top: background uncertainties as discussed in~\cref{sec:ATLAS_SS3L_Sbottom}.
		Bottom: background uncertainties taken to be 5\%. The $z$ axis shows the signal significance values.
	}
	\label{fig:Limits_13TeV_ATLAS}
\end{figure}

The projected exclusion limits obtained using the signal region that provides the best signal significance $Z$ for each mass point are presented in \cref{fig:Limits_13TeV_ATLAS}, in the $\sbottomone$ - $\ninoone$ mass plane. These limits are shown for $\sqrt{s} = 13$~TeV, with total integrated luminosities of 139~fb$^{-1}$, 300~fb$^{-1}$, and 3000~fb$^{-1}$. The red dashed lines represent the $\pm 1\sigma$ uncertainty on the number of signal events. Compared to the exclusion limits from the ATLAS Ref.~\citen{ATLAS:2019fag} analysis, the limits presented in this paper are slightly better, likely due to the simplified approach used to account for various sources of uncertainties associated with the background and signal. This is further illustrated in the three bottom-plots of \cref{fig:Limits_13TeV_ATLAS}, where an overall uncertainty value of 5\% is considered -- a value considerably smaller than the uncertainties discussed in \cref{sec:ATLAS_SS3L_Sbottom}.

\cref{fig:Limits_13TeV_ATLAS}-top shows that with 300~fb$^{-1}$, the exclusion power for supersymmetric particle masses could increase by around 100~GeV, while with 3000~fb$^{-1}$, it could increase by 200~GeV. \cref{fig:Limits_13TeV_ATLAS}-bottom indicates that the exclusion mass limits could be improved even further by reducing the uncertainties on the background estimates.

\begin{figure}[!th]
	\begin{center}\hspace{-0.7cm}
		\includegraphics[width=0.36\columnwidth]{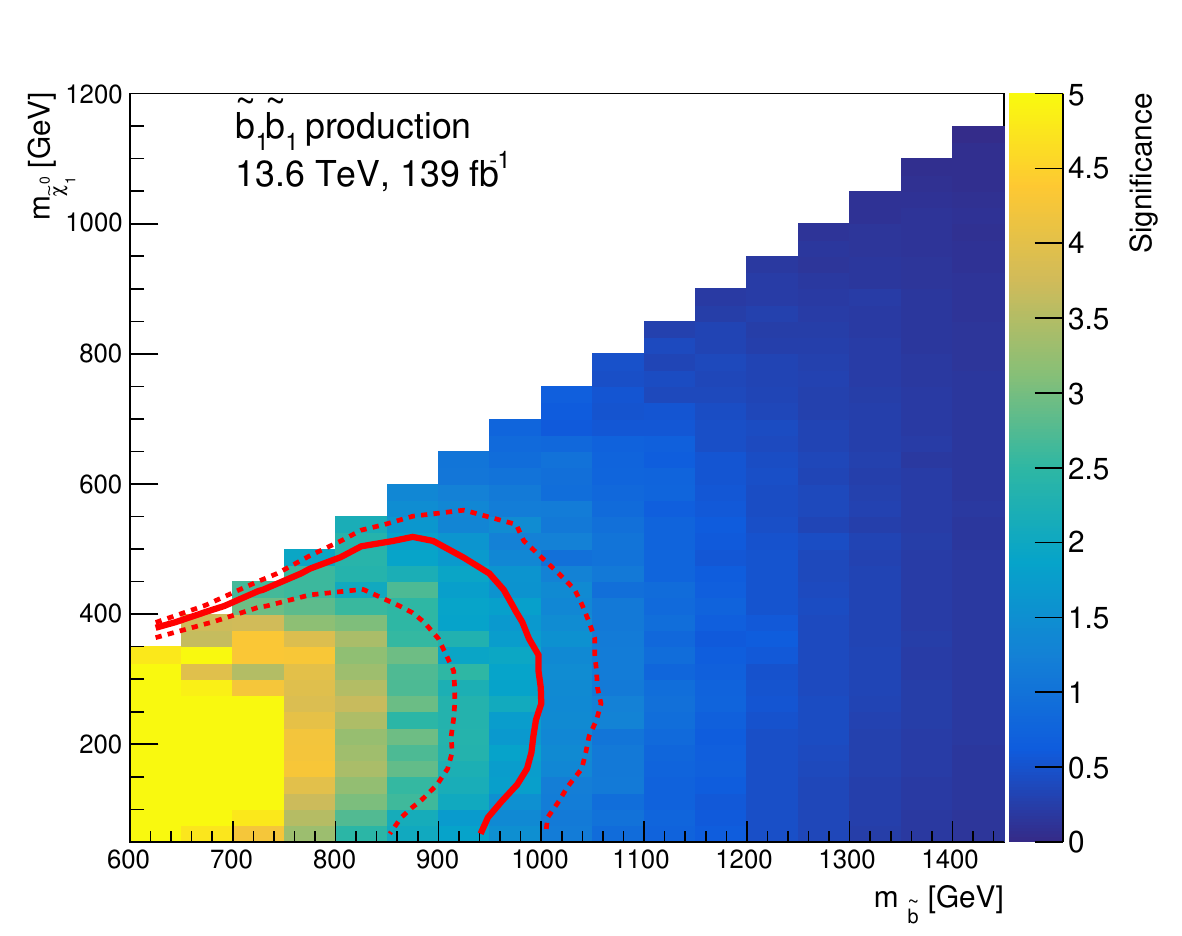}\hspace{-0.35cm}
		\includegraphics[width=0.36\columnwidth]{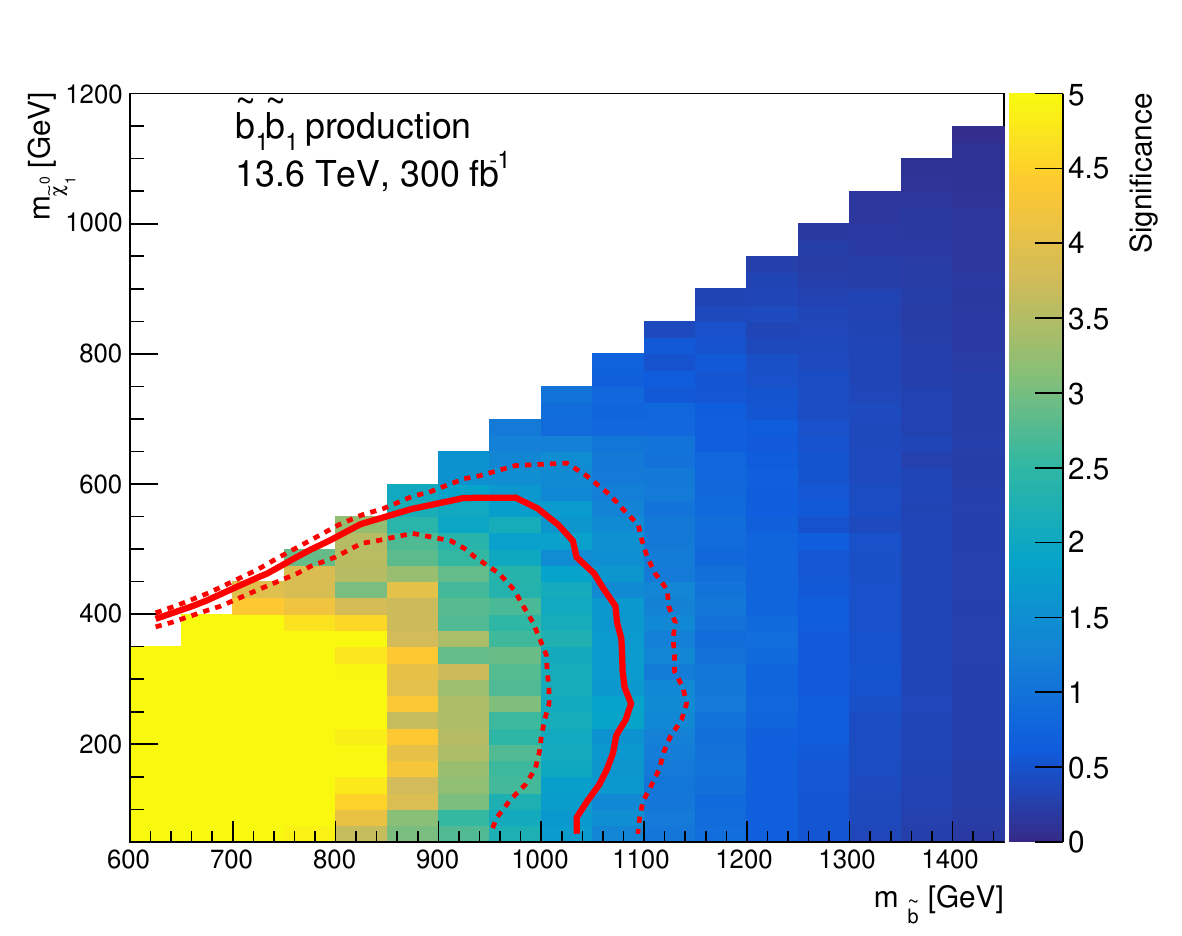}\hspace{-0.35cm}
		\includegraphics[width=0.36\columnwidth]{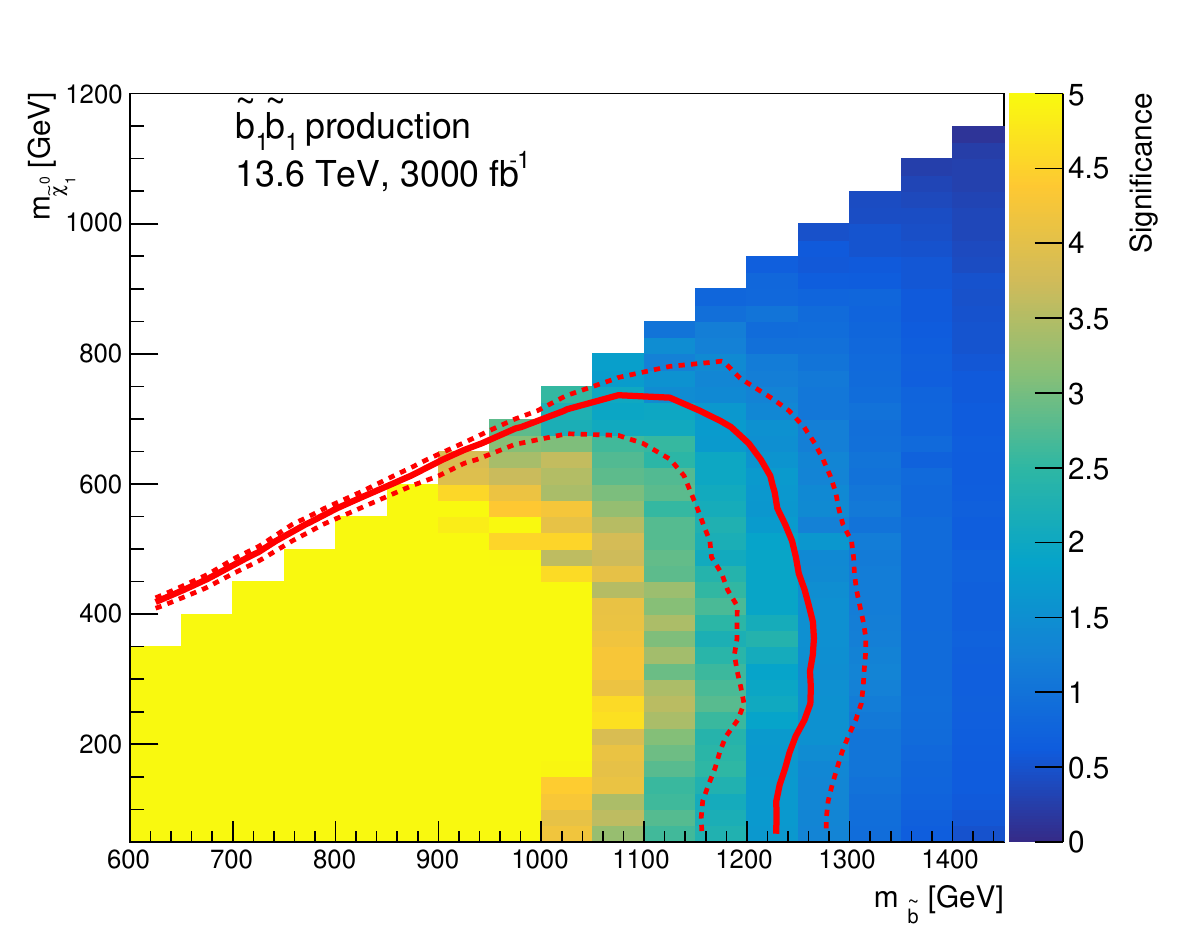}\hspace{-0.7cm}
	\end{center}
	\caption{
		Projected exclusion mass limits obtained with the signal region that gives the best signal significance $Z$ for each mass point. Projections are shown for $\sqrt s = 13.6$~TeV, and total integrated luminosities of 139~fb$^{-1}$ (left), 300~fb$^{-1}$ (middle), and 3000~fb$^{-1}$ (right). 
		The dashed lines correspond to the $\pm 1\sigma$ uncertainty on the signal event yield. The $z$ axis shows the signal significance values.
	}
	\label{fig:Limits_13p6TeV_ATLAS}
\end{figure}

The projected exclusion mass limits for $\sqrt{s} = 13.6$~TeV with total integrated luminosities of 139~fb$^{-1}$, 300~fb$^{-1}$, and 3000~fb$^{-1}$ are shown in \cref{fig:Limits_13p6TeV_ATLAS}. As expected, the increase in sensitivity due to the rise in center-of-mass energy from 13~TeV to 13.6~TeV is modest, around 40~GeV. This is because the increase in the production cross section is minimal: $0.15 \times 10^{-1}$~pb versus $0.20 \times 10^{-1}$~pb for a $\sbottomone$ mass of 900~GeV~\cite{ATLAS:XSec}. As for the $\sqrt s = 13$~TeV results, a notable gain is observed due to the expected increase in luminosity. For instance, the ATLAS Ref.~\citen{ATLAS:2019fag} analysis signal regions, without any modifications, could be sensitive in the boosted region to $\sbottomone$ masses up to 1050~GeV for a luminosity of 300~fb$^{-1}$. In the very compressed region, where weaker limits are obtained, $\sbottomone$ masses up to 850~GeV could be studied. In the less compressed region, the sensitivity ranges from 950 to 980~GeV.

With dedicated signal region optimization, the LHC Run-3 analysis sensitivity could be further enhanced. For example, to increase sensitivity in the compressed region, leading two leptons with $\pt < 20$~GeV could be considered, as well as retaining events with lower \meff and \met -- see also \cref{fig:Distrib}. This selection could increase the fake/non-prompt lepton background, but machine learning techniques could be employed to ensure sensitive signal regions. In the very boosted region, a selection with a softer requirement on the $\met$ over $\meff$ ratio could help. Binned signal regions, instead of or in addition to machine learning techniques, could further improve the sensitivity of the analysis across the entire phase space. As the four lepton regions usually have low SM background, maybe such a signal region will also help.

\begin{figure}[!th]
	\begin{center}\hspace{-0.7cm}
		\includegraphics[width=0.36\columnwidth]{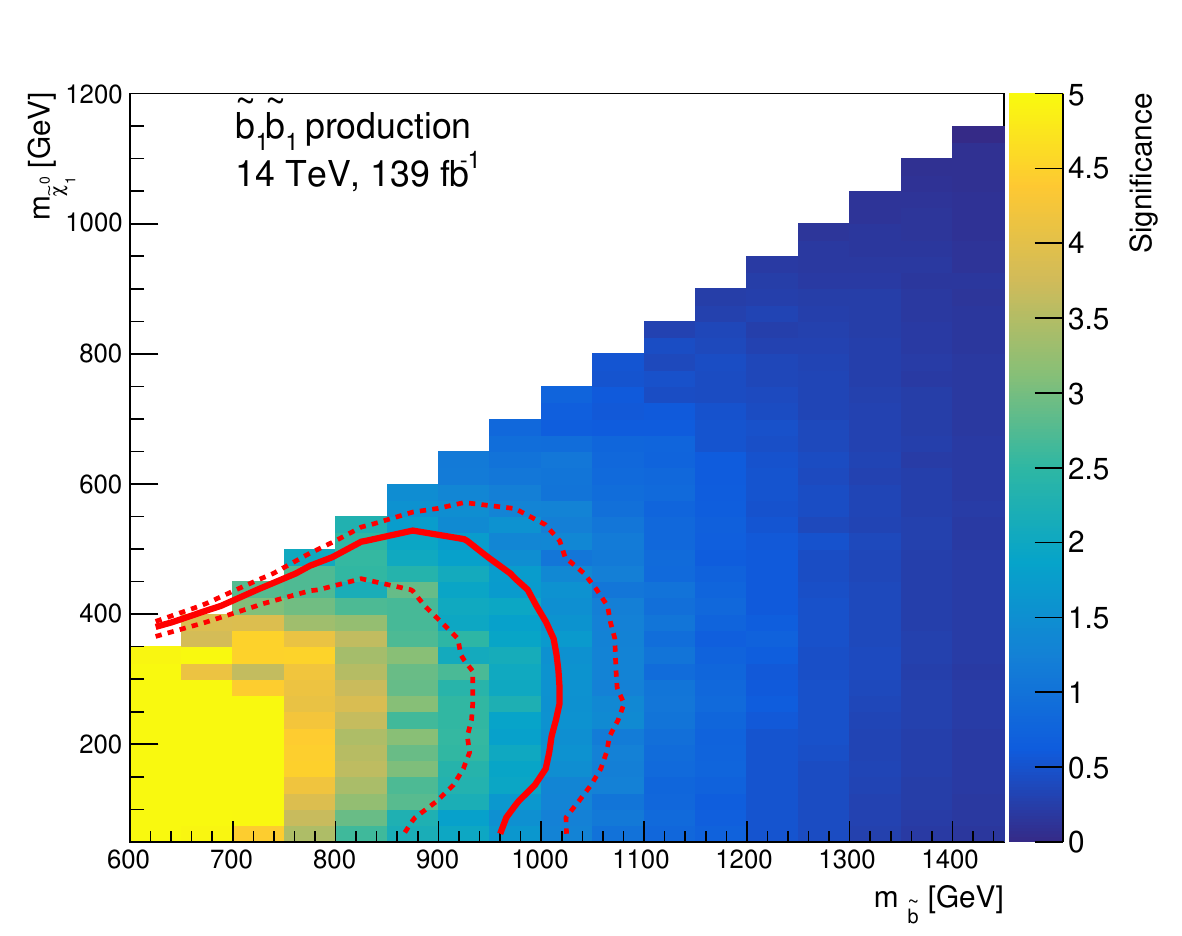}\hspace{-0.35cm}
		\includegraphics[width=0.36\columnwidth]{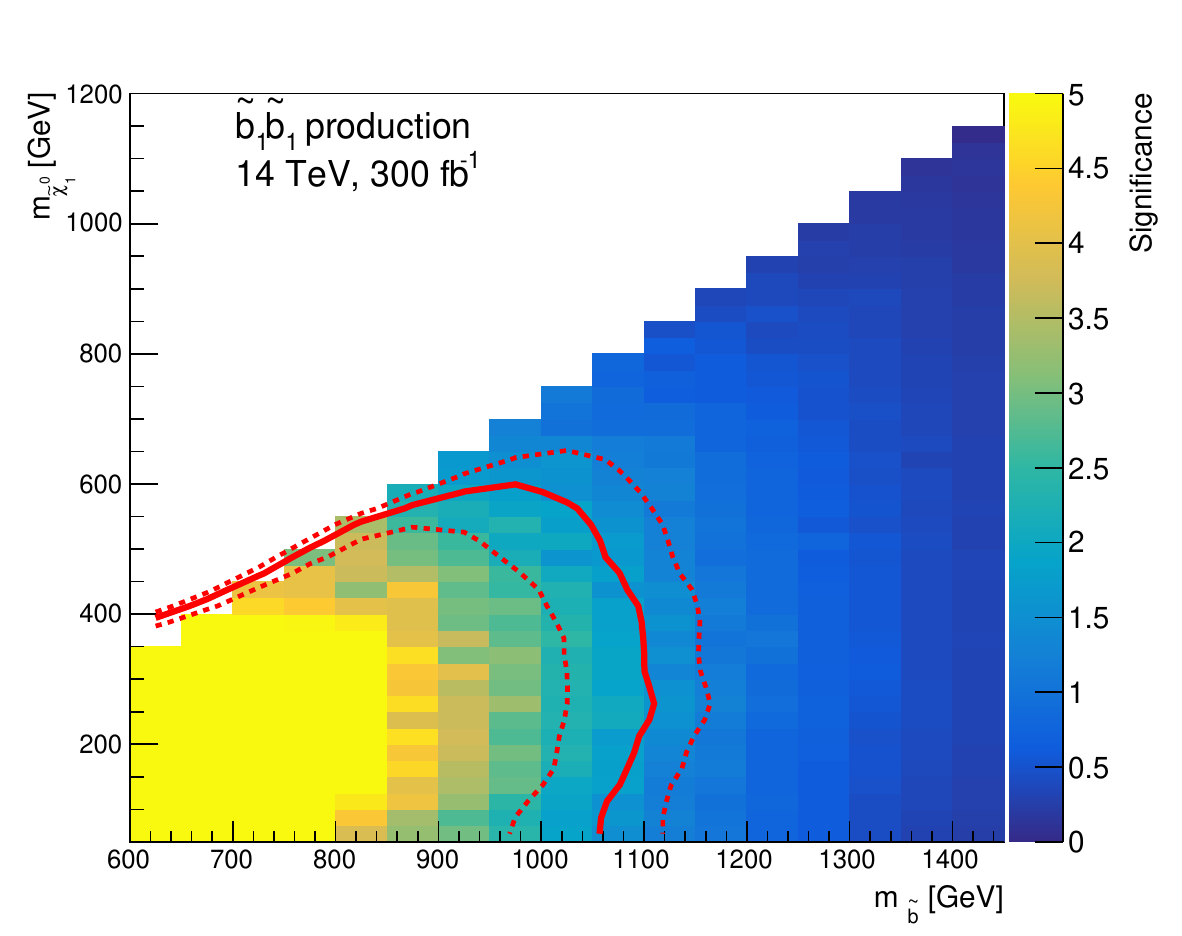}\hspace{-0.35cm}
		\includegraphics[width=0.36\columnwidth]{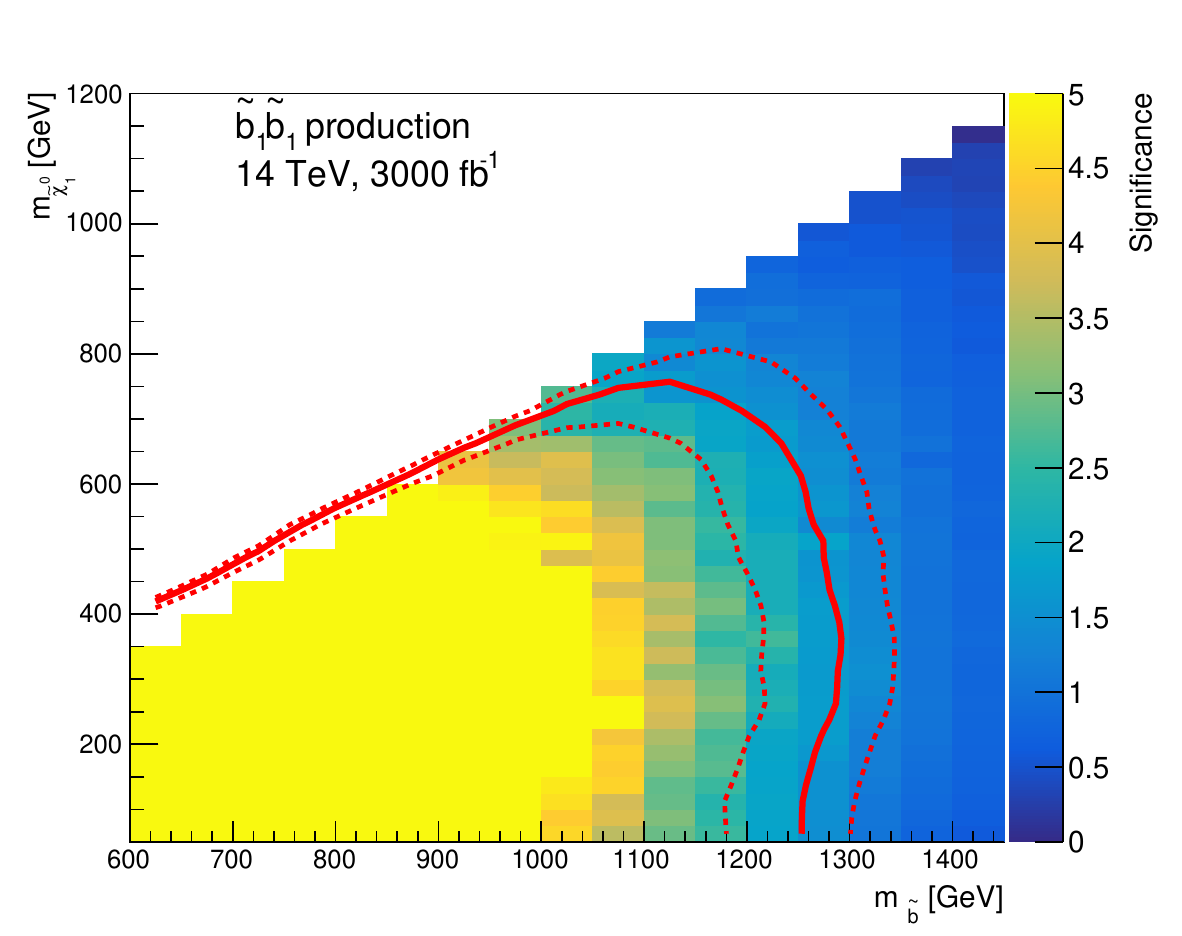}\hspace{-0.7cm}
	\end{center}
	\caption{
		Projected exclusion mass limits obtained with the signal region that gives the best signal significance $Z$ for each mass point. Projections are shown for $\sqrt s = 14$~TeV, and total integrated luminosities of 139~fb$^{-1}$ (left), 300~fb$^{-1}$ (middle), and 3000~fb$^{-1}$ (right). 
		The dashed lines correspond to the $\pm 1\sigma$ uncertainty on the signal event yield. The $z$ axis shows the signal significance values.
	}
	\label{fig:Limits_14TeV_ATLAS}
\end{figure}

Finally, \cref{fig:Limits_14TeV_ATLAS} shows the projected exclusion mass limits for $\sqrt{s} = 14$~TeV, obtained using the signal region that provides the best signal significance $Z$ for each mass point. The considered total integrated luminosities are again 139~fb$^{-1}$, 300~fb$^{-1}$, and 3000~fb$^{-1}$. To avoid long generation times, these results are derived from the \sbsbModel $\sqrt{s} = 13.6$~TeV event samples, but normalized to $\sqrt{s} = 14$~TeV production cross-sections. This is a reasonable approach, as confirmed with the 13~TeV and 13.6~TeV samples. 

Compared to 13~TeV, the analysis sensitivity could generally increase by 100~GeV, without any changes. It is interesting to observe the $Z$ values between 950~GeV (the maximum excluded by the 13~TeV limits) and 1050~GeV for the $\tilde{b}_1$ mass: they are around 5 or greater than 5. As discussed in \cref{sec:intro}, such values indicate that the considered signal process could be discovered. Thus, at the HL-LHC if the \sbsbModel model with \chinoonepm in the decay chain is true, such signals could be observed in the data. Naturally, both the discovery and exclusion potential can be improved by dedicated signal region optimization studies, as mentioned earlier.

\section{Conclusions}	
The paper presents the experimental search potential for the sbottom pair production model in an R-parity conserving scenario at the LHC Run-3 and HL-LHC, with 100\% BR one-step decay via a chargino, $\tilde{b}_1 \to t \tilde{\chi}_1^\pm$. The chargino decays to a $W$ boson and a top quark, $\tilde{\chi}_1^\pm \to W \tilde{\chi}_1^0$, again with a 100\% BR. The studies are based on ATLAS Ref.~\citen{ATLAS:2019fag}, and the final results are presented as projected exclusion limits in the $\tilde{b}_1$ - $\tilde{\chi}_1^0$ mass plane. These limits are obtained with Rpc2L1b and Rpc2L2b ATLAS Ref.~\citen{ATLAS:2019fag} signal regions defined with same-sign leptons, for three center of mass energies (13~TeV, 13.6~TeV, and 14~TeV) and three scenarios of the total integrated luminosity (139~fb$^{-1}$, 300~fb$^{-1}$, and 3000~fb$^{-1}$). Dedicated \texttt{MadGraph}+\texttt{Pythia} MC signal samples are generated, and the \texttt{DELPHES} framework is used for the ATLAS detector simulation.

It was found that, at the LHC Run-3 (13.6~TeV), the exclusion limits in the boosted could reach $\tilde{b}_1$ masses up to 1050~GeV for a luminosity of 300~fb$^{-1}$, while in the very compressed region, $\tilde{b}_1$ masses up to 850~GeV could be studied. In the less compressed region, the sensitivity is found to range from 950 to 980~GeV. These results can be further improved by optimizing the LHC Run-2 signal regions.

At the HL-LHC (14~TeV), the LHC Run-2 exclusion mass limits could generally increase by 100~GeV, without any changes to the Rpc2L1b and Rpc2L2b signal regions. The signal significance $Z$ results obtained showed that the discovery potential also significantly increased in the 950~GeV to 1050~GeV mass interval. Here too, both the discovery and exclusion potential can be increased with a re-optimization of the signal regions. Considering softer leptons, and even a channel with 4 leptons and multiple $b$-tagged jets, might further improve the analysis coverage. Machine learning techniques or binned signal regions could also enhance the sensitivity.

Both at LHC Run-3 and HL-LHC, the main sources of uncertainties are expected to be theoretical and experimental, with a significant decrease in statistical uncertainty. Additionally, the modeling uncertainties associated with the main SM backgrounds are expected to contribute less, as the large amount of data will allow for the definition of control regions to constrain these SM processes. Regarding experimental uncertainties, Ref.~\citen{ATLAS:2019mfr} shows significant improvements expected in the reconstruction, identification, and isolation of ATLAS objects, mainly thanks to the new ITk inner tracking detector.

The various upgrades of the ATLAS and CMS detectors for LHC Run-3 and HL-LHC, along with improvements in $b$ and $c$ jet tagging, and electron and muon reconstruction and identification using machine learning techniques, are opening the gates to an exciting period for new physics searches, and ideally discoveries. Even if supersymmetric particles have not yet been discovered, the prospects for SUSY searches remain promising. The increased luminosity, the advanced detection capabilities and the improvements in the analysis strategy will allow us to probe deeper into uncharted territories of the parameter space, potentially uncovering subtle signals of new physics. This era represents a significant step forward in our quest to understand the fundamental nature of the universe, offering the possibility of groundbreaking discoveries that could reshape our theoretical frameworks and guide future research directions.

\section*{Acknowledgments}
This work received support from IFIN-HH under the Contract ATLAS CERN-RO with the Romanian MCID / IFA.


\bibliographystyle{utphys}
\bibliography{MyBibliography.bib}

\end{document}